\documentclass[aps,prl,twocolumn,groupedaddress,nofootinbib,notitlepage,showpacs,floatfix]{revtex4-1}

\usepackage{graphicx,graphics,epsfig,subfigure,times,bm,bbm,amssymb,amsmath,amsfonts,amsthm,mathrsfs,MnSymbol}
\usepackage[pdfstartview=FitH]{hyperref}
\hypersetup{
    colorlinks=true,       
    linkcolor=red,          
    citecolor=magenta,        
    filecolor=magenta,      
    urlcolor=cyan,           
    runcolor=cyan
}
\usepackage{epstopdf}
\usepackage[pdftex]{color}
\usepackage{hypernat}
\usepackage{braket}
\usepackage{enumerate}
\usepackage[normalem]{ulem}
\usepackage{multirow}

\def\sx{\sigma^x}
\def\sy{\sigma^y}
\def\sz{\sigma^z}

\begin{document}

\title{Robust Quantum Control for Adiabatic Quantum Computation}
\author{Gregory Quiroz}
\affiliation{The Johns Hopkins University Applied Physics Laboratory, Laurel, Maryland, 20723, USA}

\begin{abstract}
Properly designed   
control has been shown to be particularly advantageous for improving AQC  
accuracy and time complexity scaling. Here, an \emph{in situ} quantum control optimization protocol is developed to indirectly  
optimize state fidelity without knowledge of the instantaneous spectral gap or the computational  
solution. The protocol is shown to converge to analytically-derived time-optimal controls for Grover's search algorithm.  
Furthermore, the protocol is utilized to explore optimized  
control trajectories for the Maximum 2-bit Satisifiability (MAX 2-SAT)  
problem, where appreciable improvement in fidelity and the minimum spectral gap over a linear schedule is observed. The approach is also shown to be robust against system model uncertainties (unitary control errors). This method is designed to enable robust control optimization on existing quantum annealing hardware and future AQC processors.
\end{abstract}

\maketitle

\emph{Introduction.}--Adiabatic quantum computation (AQC) utilizes controlled adiabatic evolution of a many-body quantum system to implement a quantum algorithm. The quantum system is described by a Hamiltonian $H_{ad}[\textbf{x}(t)]=\sum^L_{l=1}x_l(t)H_l$, where the ground state of the initial Hamiltonian $H_0=H_{ad}[\textbf{x}(0)]$ is assumed to be easily prepared and the ground state of the problem Hamiltonian $H_P=H_{ad}[\textbf{x}(T)]$ represents the solution to the computational problem~\cite{FarhiAQC:00,Farhi:01,AlbashAQC:16}. The system evolution is dictated by the control schedules $\textbf{x}(t)=\{x_l(t)\}$, which effectively controls the amplitude of each non-commuting, linearly independent, primitive Hamiltonian $H_l$.

The accuracy of AQC is determined by the adiabatic theorem, which asserts that the system will remain in an instantaneous eigenstate of $H_{ad}(t)$ provided the dynamics are sufficiently slow. In the noise-free case, the adiabatic theorem yields the rigorous bound on the trace-norm distance~\cite{comment-TND} between $\ket{\Phi_0(t)}$, the instantaneous ground state of $H_{ad}(t)$, and the time-evolved state $\ket{\psi(t)}$: $D[\ket{\Phi_0(T)},\ket{\psi(T)}]\lesssim q^a$, provided
\begin{equation} 
T\gtrsim \frac{a}{q}\frac{\max_{s\in\{0,1\}}\|\frac{d}{ds}H_{ad}\|^{b-1}}{\Delta^{b}_{\min}},
\end{equation}
where $s=t/T$ is the normalized time, $\|A\|$ denotes the operator norm, and $\Delta_{\min}$ is the \emph{minimum spectral gap} between the instantaneous ground state and first excited state of $H_{ad}(t)$. The parameter $q\in(0,1)$, while the integer exponents $a$ and $b$ depend upon the differentiability and analyticity properties of $H_{ad}(t)$ and the boundary conditions satisfied by its derivatives~\cite{Jansen:07,lidar:102106,Wiebe:12}. 

The adiabatic theorem has been the basis for a number of studies focused on properly designing $\textbf{x}(t)$ to minimize the adiabatic error $D$ and reduce the lower bound on $T$ by modifying $\Delta_{\min}$ via a local adiabatic condition that seeks to minimize ground state transitions $\forall t$~\cite{GroverRC:02}. These approaches have employed variational time-optimal strategies~\cite{RKHLZ:09}, optimal control theory~\cite{BrifAQC:14}, and convex optimization~\cite{ZengPathOpt:16} that exploit accurate system models and either knowledge of the computational solution, i.e. $\ket{\Phi_0(T)}$, or the instantaneous spectral gap $\Delta(s)$. In this work, an optimization technique referred to as Closed-Loop Optimized Adiabatic Quantum Control (CLOAQC) is developed to indirectly optimize the adiabatic error using only the time-evolved system state at $t=T$ measured in the computational basis. The method is shown to converge towards known time-optimal solutions for Grover's search algorithm (GSA)~\cite{GroverRC:02,RKHLZ:09} and substantially improve adiabatic error and enhance $\Delta_{\min}$ for the MAX 2-SAT problem relative to a linear schedule. The protocol is shown to exhibit robustness to unitary control errors and it is argued that due to the form of the objective function, the method can be readily extended to more generic noise models.

\emph{CLOAQC protocol.} -- Closed-loop quantum control learning is an iterative optimization method that relies on information from previous experiments to update control parameters and effectively optimize system performance with respect to a given objective function. In the case of AQC control optimization, the learning procedure includes three main steps: (1) the generation of a set of control parameters, (2) the implementation of the AQC algorithm and subsequent quantification of performance, and (3) a learning algorithm that incorporates prior performance information to provide updated control parameters. Observing the impact of varying the control parameters via prior experiments is a key aspect of closed-loop learning that affords inherent robustness to system uncertainty. It is exploited here to consider a quantum processor designed to implement quantum annealing or more generally, AQC in a blackbox framework where one has limited knowledge of the intrinsic noise processes, systematic errors, and the underlying structure of the energy spectrum of $H_{ad}(t)$. To this end, it is assumed that one only has knowledge of the Hamiltonian one \emph{believes} is being implemented on the hardware and the state of the system at the \emph{end} of the computation measured in the computational basis. These assumptions fit well within the confines of currently available quantum annealing based hardware, such as the D-Wave processor~\cite{Dwave}, and future AQC processors. 

A function space is used to parametrize the control functions for each constituent Hamiltonian $H_i$ in $H_{ad}[\textbf{x}(s)]$. Each control function is defined as
\begin{equation}
x_i(s) = \sum^{d+1}_{j=1}\alpha_{ij}\phi_j(s),
\end{equation}
where $\alpha_{ij}$ denote weights for each of the $j$th basis functions $\phi_j(s)$. Thus, the optimization parameter space is defined by $\Lambda\equiv\{\alpha_{ij}\}^{L,d+1}_{i,j=1}$, which includes the weights for the $d$ basis elements in the function expansion for all $L$ controls. Note that the function representation has advantages over piecewise control in that the parameter space is drastically reduced and variations in control parameters result in global rather than local changes in control functions. Here, the polynomial basis $\phi_k(s)=s^{k-1}$ is chosen for simplicity; however, one can readily consider any alternative. Intrinsically band-limited functions such as the discrete prolate spheriodal sequences (DPSS) may be attractive for imposing intrinsic bandwidth constraints on control profiles~\cite{Lucarelli:16}.

Traditional state fidelity metrics, such as the trace-norm distance, require knowledge of the time evolved state at time $T$, $\ket{\psi(T,\Lambda)}$, and the target ground state of $H_P$, $\ket{\Phi_0(T)}$. While one may estimate $\ket{\psi(T,\Lambda)}$ by sampling the AQC algorithm, knowledge of $\ket{\Phi_0(T)}$ implies knowledge of the computational solution. One may envision control protocols that exploit partial or approximate solutions obtained from classical algorithms, however, the focus of this work is the case where the computational solution is unknown and additional classical preprocessing (i.e. approximate solution optimization algorithm) can be avoided. 


State fidelity metrics are circumvented by employing the energy of the system with respect to $H_P$, $E(\Lambda) = \braket{\psi(T,\Lambda)| H_P | \psi (T,\Lambda)}$ to evaluate the performance of a given set of control parameters $\Lambda$. The expectation value is estimated by accumulating samples of the final state measured in the computational basis $\{\ket{\tilde{\psi}_i(T,\Lambda)}\}^{M}_{i=1}$ from $M$ implementations of the AQC algorithm; this estimate is denoted as $\hat{E}(\Lambda)$. Note that $E(\Lambda)$ serves as a viable surrogate objective function that achieves a minimum value when $\ket{\psi(T,\Lambda)} = \ket{\Phi_0(T)}$; see Ref.~\cite{comment-energy} regarding variants of $E(\Lambda)$ previously considered.

Due to sampling statistics, function calls to the $\hat{E}(\Lambda)$ are stochastic. Therefore, a stochastic optimization technique must be employed to perform the control optimization. Simultaneous Perturbative Stochastic Approximation (SPSA), an iterative, gradient-based optimization technique that requires two function calls per iteration to estimate the gradient is selected for this task~\cite{SpallSPSA:92}. Note that SPSA has been previously used for quantum information applications~\cite{FerrieSGQT:14,FerrieQC:15,Granade:15}. For each iteration $k$, the first step is to generate a random search vector $\Delta_k$, where each of the $i=1,2,\ldots, |\Lambda|$ elements are Bernoulli distributed variables, i.e. $\Delta_{kj}=\pm 1$. The estimated gradient is 
\begin{equation}
g_k = \frac{\hat{E}(\Lambda_k+\beta_k \Delta_k) - \hat{E}(\Lambda_k - \beta_k \Delta_k)}{2 \beta_k},
\label{eq:gradient}
\end{equation}
where control parameter update is given by 
\begin{equation}
\Lambda_{k+1} = \Lambda_k + \alpha_k [g_k + \lambda_{ad}\nabla J_{ad} (\Lambda_k)]
\end{equation}
and $\lambda_{ad}=0.005$ in all subsequent simulations. The update includes an additional analytically calculated gradient for the objective function
\begin{equation}
J_{ad}=\sum_{\mu\in \textbf{x}}\int^{1}_{0}\|\dot{\mu}(s)\|ds,
\end{equation}
which seeks to enforce adiabaticity by minimizing the derivative of each control field over the total time interval \cite{BrifAQC:14}.  

The functions $\alpha_k$ and $\beta_k$ are convergence parameters and typically defined as 
\begin{equation}
\alpha_k = \frac{\alpha_0}{(k+1+R)^\delta}, \quad \beta_k = \frac{\beta_0}{(k+1)^\zeta},
\end{equation}
where $\alpha_0$, $\beta_0$, $R$, are chosen following the procedure outlined in Ref.~\cite{SpallSPSAImp:98}. The remaining parameters $\delta=0.602$ and $\zeta=0.101$ are typically good values \cite{SpallSPSA:92} and appear to be good choices for the problem presented in this work. It is also found that the asymptotically optimal values $\delta=1$ and $\zeta=1/6$ yield similar results~\cite{SadeghSPSA:98}.

\emph{Convergence and performance.} -- CLOAQC requires a total number of $2MK$ experiments, where $K$ is the number of iterations of the algorithm. The factor of two arises from SPSA's finite difference gradient [Eq.~(\ref{eq:gradient})] estimation, which differs from the $2d$ estimates required for standard finite difference gradient techniques.

SPSA convergence analyses~\cite{SpallSPSA:92,SadeghSPSA:98} indicate a reduction in $E(\Lambda)$ at a rate of $\mathcal{O}(1/k^\gamma)$. This scaling is also shown to hold for the adiabatic error $D$. The exponent $\gamma$ is highly problem dependent, but asymptotic results indicate $\gamma\approx 1$ to first order.

\emph{Grover's search algorithm.} -- The efficacy of CLOAQC is explored via Grover's search algorithm (GSA) for the identification of a marked element in an unsorted database of $N$ elements~\cite{Grover:97}. GSA requires a minimum of $\mathcal{O}(\sqrt{N})$ oracle queries to identify the marked element, a quadratic improvement over the best possible classical algorithm~\cite{Bennett:97}. Recast in the language of AQC~\cite{GroverRC:02,RKHLZ:09}, Grover's algorithm is generically defined by an $n$-qubit Hamiltonian
\begin{equation}
H_{G}(s) = x_1(s) [I-\ket{+}\bra{+}] + x_2(s) [I-\ket{m}\bra{m}], 
\end{equation}
where $x_{1,2}(s)$ are the control functions, $I$ is the identity operator, $\ket{+}$ represents the uniform superposition over all $N=2^n$ computational basis states, and $\ket{m}$ is the marked state. Time-optimal controls can be designed such that $\Delta_{\min}\sim\mathcal{O}(1/\sqrt{N})$, and the total runtime required to reach the ground state $\ket{\Phi_0(T)}=\ket{m}$ is $T\sim\mathcal{O}(\sqrt{N})$; thus, achieving the well-known quadratic speedup~\cite{GroverRC:02,RKHLZ:09}.

The optimized control functions obtained from CLOAQC are compared to the time-optimal GSA controls for one independent control (IC) [$x_1(s)=1-x_2(s)$] and two ICs, where $x_1(s)$ and $x_2(s)$ are linearly independent. The CLOAQC algorithm is initialized such that $\Lambda_0$ describes a linear ramping control schedule. The boundary conditions $x_1(0)=x_2(1)=1$ and $x_1(1)=x_2(0)=0$ are enforced on the control profiles throughout the optimization procedure. Each control function is expanded into five basis functions, and thus, the control parameter space is described by five and ten parameters in the one and two IC case, respectively. The total runtime $T$ is chosen so that there is a 40\% probability of being in the ground state of $H_P$; see Appendix for further details on runtime specifications.

\begin{figure}[t]
\includegraphics[width=\columnwidth, height=0.7\columnwidth]{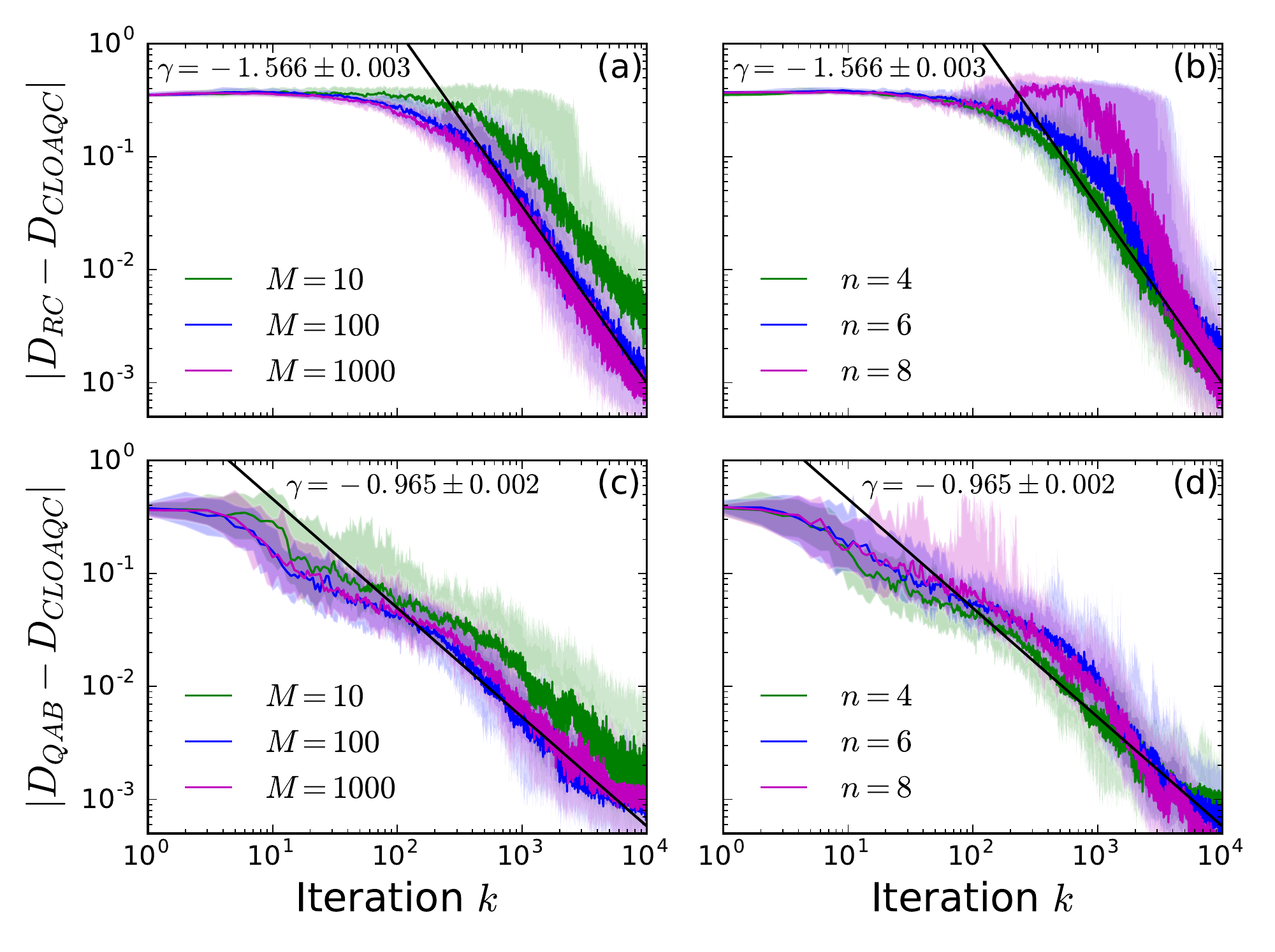}
\caption{Relative difference between trace-norm distances for time-optimal GSA controls and CLOAQC for one and two control degrees of freedom. Comparison illustrates the dependence of CLOAQC performance on the number of algorithm implementations $M$ and number of qubits $n$.}
\label{fig:grover}
\end{figure}

A comparison of performance indicates a convergences in CLOAQC solutions toward the time-optimal GSA solutions with increasing iteration. 
In Figure~\ref{fig:grover}, CLOAQC is compared to the Roland-Cerf (RC)~\cite{GroverRC:02} and the quantum adiabatic Brachistochrone (QAB)~\cite{RKHLZ:09} GSA solutions for various values of sampling parameter $M$ and number of qubits $n$ using 100 realizations of CLOAQC. 
The top and bottom rows illustrate the relative difference in adiabatic error $D$ between CLOAQC and RC and QAB, respectively, as a function of iteration $k$. Median CLOAQC performance is denoted by the solid colors, while shaded region denotes the interquartile range. Although both RC and QAB solutions require spectral gap $\Delta(s)$, CLOAQC is capable of converging toward the equivalent time-optimal solutions using only the state of the system at the end of the computation. In panels (a) and (c), the convergence rate of CLOAQC is shown to be approximately independent of the sampling parameter $M$. The most considerable improvement in performance is observed between $M=10$ and $M=100$ for both control scenarios. CLOAQC convergence does not convey a compelling dependence on $n$, provided the runtime is adjusted to maintain the target ground state sampling at initialization; see Fig.~\ref{fig:grover}(b) and (d). Note that each panel includes a fit for $n=4$ with $M=100$, along with the corresponding convergence parameter $\gamma$. As expected, $\gamma$ is strongly dependent upon the number of optimization variables. 

CLOAQC optimized control profiles are qualitatively similar to the time-optimal GSA solutions. In particular, for one IC, optimized controls closely resemble the RC solution, while deviations from the QAB control profiles are more significant for two ICs. Higher order polynomial expansions are needed to more accurately reproduce the QAB profiles. Note that despite such a distinction between CLOAQC and QAB paths, their adiabatic errors only differ by less than $10^{-3}$ for the $n$ values considered here. See the appendix for a more detailed discussion.


\emph{Optimized controls for MAX 2-SAT.} --
The utility of CLOAQC is further illustrated via the Maximum 2-bit Satisfiability (MAX 2-SAT) problem, where optimized controls are shown to offer substantial reductions in adiabatic error and amplification of $\Delta_{\min}$ relative to a linear control schedule. MAX 2-SAT offers variability in the energy spectrum, including location and magnitude of $\Delta_{\min}$ that is highly problem instance dependent. By focusing on unique satisfying assignment (USA) instances (i.e., instances with non-degenerate ground state manifolds), CLOAQC's ability to maintain adiabaticity and effectively navigate the ground state manifold without knowledge of the specifications of $\Delta_{\min}$ is demonstrated. CLOAQC's ability to enhance adiabatic error is supplemented by gap amplifications that generally become more pronounced with increasing ICs. 

CLOAQC performance is assessed with respect to an ensemble of 100 USA 2-SAT instances. Each 2-sat instance is a logical AND of $M_c$ clauses, where each clause $C_j$ itself is a logical OR of exactly two Boolean variables from the set $\{x_i\}^n_{i=1}$. Each 2-SAT problem Hamiltonian is constructed by associating the binary values of each Boolean variable $x_j$ with the $\pm1$ eigenstates of the Pauli spin operator $\sz_j$ for the $j$th qubit and summing the $M_c$ clause Hamiltonians~\cite{FarhiAQC:00}. After rescaling and dropping the constant term, the resulting problem Hamiltonian is
\begin{equation}
H_P = \sum_{j} h_{j}\sz_{j} + \sum_{i,j} J_{ij}\sz_i\sz_j,
\end{equation}
where $h_j=-\sum_m v^m_j$ and $J_{ij}= \sum_m v^m_i v^m_j$ \cite{Santra2SAT:14}. The variables $v^m_j\in\{-1,0,1\}$, where $j=\{1,2,\ldots, n\}$ and $m=\{1,2,\ldots,M_c\}$ label the variables and clauses, respectively, and encode the specifications of $C_m$. Namely, if $x_j$ appears negated (unnegated) in the $m$th clause then $v^m_j=-1 (+1)$. $v^m_j=0$ for all clauses where $x_j$ does not appear. The USA ensemble is generated using the approach in Ref.~\cite{Farhi:01} and found to possess an average clause density $\hat{\alpha}=n/M_c\approx1.2$. Clause density is known to play a role in discerning problem hardness. $\hat{\alpha}$ is found to be sufficiently close to the critical clause density $\alpha_c=1$, where the most (classically) difficult MAX 2-SAT problems lay~\cite{Coppersmith2SAT:04}.





A variety of control scenarios are considered by defining the MAX-2 SAT algorithm as
\begin{equation}
H_{2S}(s) = x_1(s) H_0 + x_2(s) H_I + x_3(s) H_{P,1} + x_4(s) H_{P,2}.
\label{eq:maxsatH}
\end{equation}
The initial Hamiltonian $H_0=\sum_j \sx_j$ represents a transverse field on each qubit; thus, defining the initial ground state to be the uniform superposition state over all computational basis states. The intermediate Hamiltonian $H_I=\sum_{i\neq j}\sx_i \sx_j$ is only present for $t\in(0,T)$ and defines a non-stoquastic contribution. Non-stoquastic Hamiltonians have been studied for the MAX 2-SAT problem ~\cite{CrossonNonSto:14}, and they have been shown to benefit algorithmic performance for certain algorithms~\cite{FarhiNonSto:11,HormoziNonSto:17,NishimoriNonSto:17}. Here, their advantages in the presence of optimized control are investigated. The last two terms in Eq.~(\ref{eq:maxsatH}), $H_{P,1}=\sum_j h_j \sz_j$ and $H_{P,2}=\sum_{i,j}J_{ij}\sz_i\sz_j$, denote the 1 and 2-local terms of the problem Hamiltonian $H_P=H_{P,1}+H_{P,2}$. The number of independent controls is varied by imposing constraints on $x_i(s)$. Four scenarios are considered here: (1) one IC: $x_1(s)$ and $x_4(s)=1-x_1(s)$ are non-zero, (2) two ICs: $x_2(s)=0$, $x_3(s)=x_4(s)$, (3) three ICs: $x_3(s)=x_4(s)$, and (4) four ICs, where all control functions vary independently. In all cases, $x_1(0)=x_3(1)=x_4(1)=1$, $x_1(1)=x_3(0)=x_4(0)=0$, and $x_2(0)=x_2(1)=0$ are imposed on the control functions.

\begin{figure}[t]
\includegraphics[width=\columnwidth, height=0.55\columnwidth]{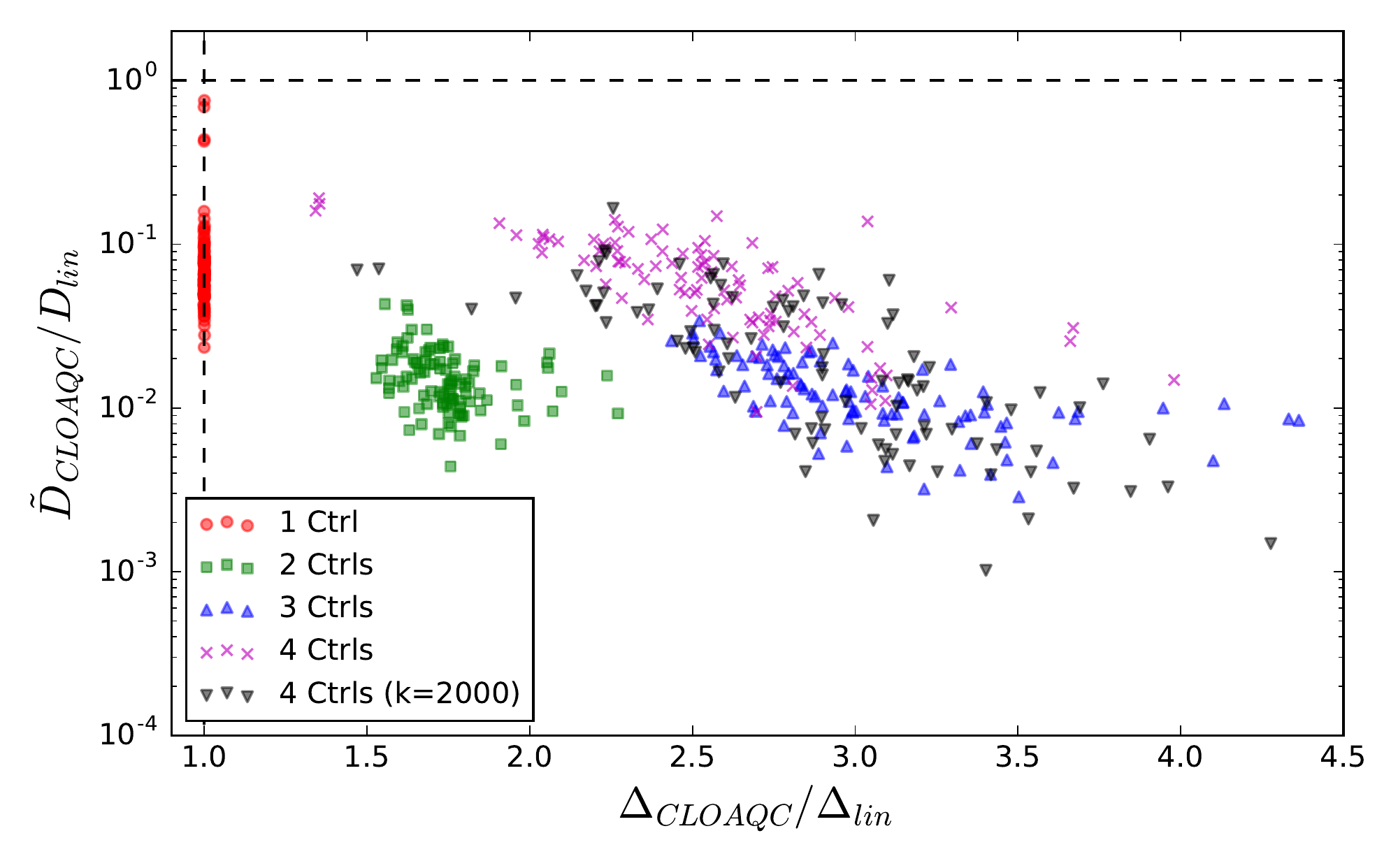}
\caption{Comparison of CLOAQC vs. a linear schedule for 100 instances of the MAX 2-SAT problem for four different control scenarios.}
\label{fig:max2sat}
\end{figure}

CLOAQC is shown to outperform a linear control schedule for the ensemble of 100 USA instances in Figure~\ref{fig:max2sat}. Median adiabatic error $\tilde{D}$ is obtained from an ensemble of 25 realizations of CLOAQC for each problem instance, with $\Delta_{CLOAQC}$ corresponding to the minimum gap for the control profile that produces $\tilde{D}$. CLOAQC is implemented for 1000 iterations with $M=100$. 

Performance features between different control scenarios are remarkably distinct. Letting $\alpha_D$ and $\alpha_\Delta$ denote the medians of $\tilde{D}_{CLOAQC}/D_{lin}$ and $\Delta_{CLOAQC}/\Delta_{lin}$ with respect to the distribution of problem instances, control scenarios are compared against each other by their median improvement in adiabatic error and gap enhancement. In the one control case, $\alpha_D\approx 0.066$ and no gap amplification is observed. A minimum of two ICs are required to achieve gap amplifications, where $\alpha_\Delta\approx 2$ and the median adiabatic error ratio improves to $\alpha_D\approx 0.011$. Further improvements in adiabatic error and minimum gap size are achievable by including $H_I$ with optimized control, however, the degree of improvement is strongly dependent upon the choice of $H_I$ and problem instance; this is consistent with previous findings~\cite{HormoziNonSto:17}. The distribution for the three IC case is fairly localized in adiabatic error and broad in gap enhancement, where the median performance ratio $\alpha_D\approx 0.011$ and $\alpha_\Delta \approx 2.97$. Note that in some cases the gap enhancement reaches as larger as approximately $4.4\times \Delta_{lin}$. While a subset of the instances do benefit from four ICs, a majority of the instances do not. The median ratio of adiabatic error is $\alpha_D\approx 0.062$ and the median gap enhancement is $\alpha_\Delta 2.538$ for the distribution of instances. The degradation in performance is due to the fixed number of CLOAQC iterations and increasing dimension of the search space. Increasing the number of iterations to $K=2000$, CLOAQC performance improves to $\alpha_D\approx 0.020$ and a modest median gap enhancement of $\alpha_\Delta \approx 2.89$. Further improvements in $D$ likely require an increase in $M$ and $K$, specifically in the high fidelity regime where achieving non-zero $\hat{E}(\Lambda)$ values may require $M\gg 1$. See appendix for analyses of alternative intermediate Hamiltonians.

\begin{figure}[t]
\centering
\includegraphics[width=0.9\columnwidth, height=0.6\columnwidth]{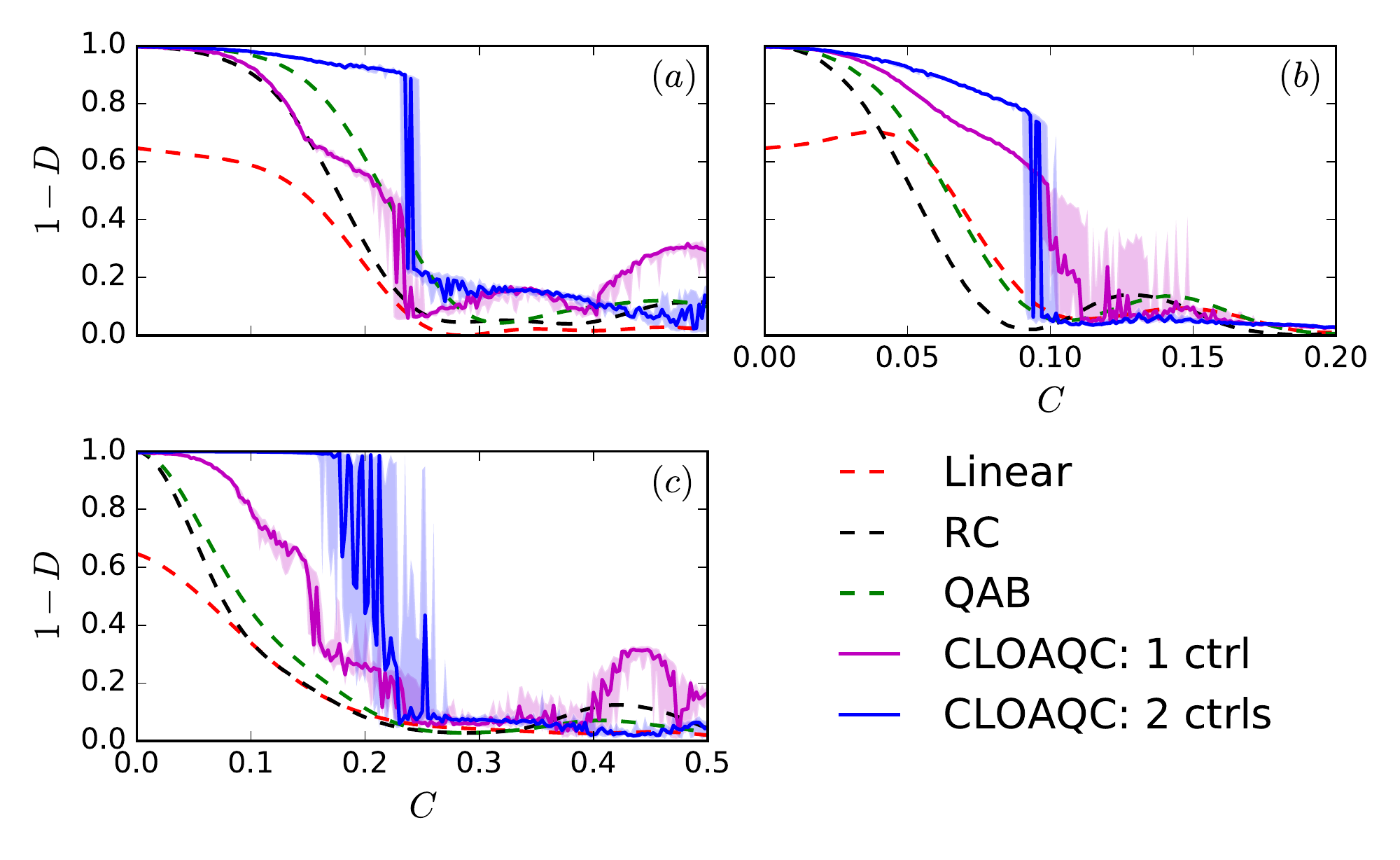}
\caption{Comparison of CLOAQC against linear, RC, and QAB control schedules for the three unitary control error models discussed in the main text. Median CLOAQC performance is shown for a distribution of 25 realizations of CLOAQC using $K=1000$ and $M=100$.}
\label{fig:gsa_hp}
\end{figure}

\emph{Robustness to noise.} -- CLOAQC possesses a degree of inherent robustness to uncertainty in $H_{ad}(s)$ due to the fact that it is a closed-loop protocol and it relies on the minimization of the average energy with respect to $H_P$, the problem Hamiltonian one wishes to encode on the AQC hardware. 
CLOAQC's robustness is assessed here by including an additive, unitary control error with 3 different types of ramping schedules. The Hamiltonian 
\begin{equation}
H^\prime_{ad}(s) = H_{ad}(s) + H_E(s)
\end{equation}
is used to describe the faulty AQC algorithm, with the additional additive term $H_E(s)=\Gamma(s) \sum^n_{i} \hat{m}_i\cdot \vec{\sigma}_i$ contributing to the deformation of the ground state manifold. $\hat{m}_i=(m_{i,x}, m_{i,y}, m_{i,z})$ is a unit vector where $m_{i,\mu}$ is generated from a zero mean normal distribution with unit standard deviation. The ramping schedule $\Gamma(s)$ is chosen in accordance with Ref.~\cite{ChildsAQC:01} and takes three forms: (a) $\Gamma(s)=C s$, (b) $\Gamma(s)=C\sin(\pi s)$, and (c) $\Gamma(s)=1/2\sin(C\pi s)$, where $C\in \mathbb{R}$. Figure~\ref{fig:gsa_hp} compares CLOAQC to a linear, RC, and QAB control profile for GSA, with each panel corresponding to the three $\Gamma(s)$ schedules, respectively. Median performance is shown for a distribution of 25 realizations of CLOAQC, using a realization of $H_E(t)$ that does $\emph{not}$ exhibit favorable recurrences in $D$ with increasing $C$~\cite{ChildsAQC:01}. CLOAQC conveys considerable improvements in adiabatic error and robustness for sufficiently small and slow-oscillating unitary control errors, most notably for ramping schedules (b) and (c), where one-control CLOAQC outperforms (two-control) QAB. CLOAQC performance exhibits a abrupt degradation in performance at critical $C$ values where the dynamics are dominated by $H_E(s)$ and insufficient sampling of the $H_P$ ground state exists.

While the focus here is unitary control errors, given the form of the objective function, CLOAQC is expected to be robust against more generic noise sources. Generically, CLOAQC can be expected to outperform a linear schedule when $\|H_{ad}(s)\|\gg \|H_E(s)\|$, where $\|H_E(s)\| < \infty$ and $\|\cdot\|$ is the operator norm, and the dynamics generated by $H_E(s)$ are sufficiently slow $\forall s$. The development of rigorous bounds and potential relaxations of this local condition on $H_E(s)$ related to CLOAQC performance in the open quantum system setting is left for future work.

\emph{Conclusions.} -- A blackbox AQC control optimization protocol (referred to as CLOAQC) is presented and shown to drastically improve algorithmic performance solely using the state of the quantum system at the end of the computation and no knowledge of the minimum spectral gap. CLOAQC is shown to converge towards grover's search algorithm's time-optimal control solutions with increasing iteration and achieve robustness to uncertainties in the Hamiltonian, specifically unitary control errors. Further exploring the utility of the approach, CLOAQC is used to optimize control profiles for the MAX 2-SAT problem and achieve improvements in computational accuracy and the size of the minimum energy gap relative to a linear control schedule. CLOAQC is designed to be conducive to current quantum annealing hardware and future AQC processors.


\emph{Acknowledgements} -- The author thanks B. D. Clader and Dennis Lucarelli for useful discussions.

%
%

\begin{thebibliography}{33}%
\makeatletter
\providecommand \@ifxundefined [1]{%
 \@ifx{#1\undefined}
}%
\providecommand \@ifnum [1]{%
 \ifnum #1\expandafter \@firstoftwo
 \else \expandafter \@secondoftwo
 \fi
}%
\providecommand \@ifx [1]{%
 \ifx #1\expandafter \@firstoftwo
 \else \expandafter \@secondoftwo
 \fi
}%
\providecommand \natexlab [1]{#1}%
\providecommand \enquote  [1]{``#1''}%
\providecommand \bibnamefont  [1]{#1}%
\providecommand \bibfnamefont [1]{#1}%
\providecommand \citenamefont [1]{#1}%
\providecommand \href@noop [0]{\@secondoftwo}%
\providecommand \href [0]{\begingroup \@sanitize@url \@href}%
\providecommand \@href[1]{\@@startlink{#1}\@@href}%
\providecommand \@@href[1]{\endgroup#1\@@endlink}%
\providecommand \@sanitize@url [0]{\catcode `\\12\catcode `\$12\catcode
  `\&12\catcode `\#12\catcode `\^12\catcode `\_12\catcode `\%12\relax}%
\providecommand \@@startlink[1]{}%
\providecommand \@@endlink[0]{}%
\providecommand \url  [0]{\begingroup\@sanitize@url \@url }%
\providecommand \@url [1]{\endgroup\@href {#1}{\urlprefix }}%
\providecommand \urlprefix  [0]{URL }%
\providecommand \Eprint [0]{\href }%
\providecommand \doibase [0]{http://dx.doi.org/}%
\providecommand \selectlanguage [0]{\@gobble}%
\providecommand \bibinfo  [0]{\@secondoftwo}%
\providecommand \bibfield  [0]{\@secondoftwo}%
\providecommand \translation [1]{[#1]}%
\providecommand \BibitemOpen [0]{}%
\providecommand \bibitemStop [0]{}%
\providecommand \bibitemNoStop [0]{.\EOS\space}%
\providecommand \EOS [0]{\spacefactor3000\relax}%
\providecommand \BibitemShut  [1]{\csname bibitem#1\endcsname}%
\let\auto@bib@innerbib\@empty
\bibitem [{\citenamefont {Farhi}\ \emph {et~al.}()\citenamefont {Farhi},
  \citenamefont {Goldstone}, \citenamefont {Gutmann},\ and\ \citenamefont
  {Sipser}}]{FarhiAQC:00}%
  \BibitemOpen
  \bibfield  {author} {\bibinfo {author} {\bibfnamefont {E.}~\bibnamefont
  {Farhi}}, \bibinfo {author} {\bibfnamefont {J.}~\bibnamefont {Goldstone}},
  \bibinfo {author} {\bibfnamefont {S.}~\bibnamefont {Gutmann}}, \ and\
  \bibinfo {author} {\bibfnamefont {M.}~\bibnamefont {Sipser}},\ }\href
  {http://arxiv.org/abs/quant-ph/0001106} {\ }\Eprint
  {http://arxiv.org/abs/quant-ph/0001106} {quant-ph/0001106} \BibitemShut
  {NoStop}%
\bibitem [{\citenamefont {Farhi}\ \emph {et~al.}(2001)\citenamefont {Farhi},
  \citenamefont {Goldstone}, \citenamefont {Gutmann}, \citenamefont {Lapan},
  \citenamefont {Lundgren},\ and\ \citenamefont {Preda}}]{Farhi:01}%
  \BibitemOpen
  \bibfield  {author} {\bibinfo {author} {\bibfnamefont {E.}~\bibnamefont
  {Farhi}}, \bibinfo {author} {\bibfnamefont {J.}~\bibnamefont {Goldstone}},
  \bibinfo {author} {\bibfnamefont {S.}~\bibnamefont {Gutmann}}, \bibinfo
  {author} {\bibfnamefont {J.}~\bibnamefont {Lapan}}, \bibinfo {author}
  {\bibfnamefont {A.}~\bibnamefont {Lundgren}}, \ and\ \bibinfo {author}
  {\bibfnamefont {D.}~\bibnamefont {Preda}},\ }\href {\doibase
  10.1126/science.1057726} {\bibfield  {journal} {\bibinfo  {journal}
  {Science}\ }\textbf {\bibinfo {volume} {292}},\ \bibinfo {pages} {472}
  (\bibinfo {year} {2001})}\BibitemShut {NoStop}%
\bibitem [{\citenamefont {Albash}\ and\ \citenamefont
  {Lidar}(2016)}]{AlbashAQC:16}%
  \BibitemOpen
  \bibfield  {author} {\bibinfo {author} {\bibfnamefont {T.}~\bibnamefont
  {Albash}}\ and\ \bibinfo {author} {\bibfnamefont {D.}~\bibnamefont {Lidar}},\
  }\href {https://arxiv.org/abs/1611.04471} {\  (\bibinfo {year} {2016})},\
  \Eprint {http://arxiv.org/abs/ArXiv:1611.04471} {ArXiv:1611.04471}
  \BibitemShut {NoStop}%
\bibitem [{com({\natexlab{a}})}]{comment-TND}%
  \BibitemOpen
  \href@noop {} {} \bibinfo {note} {\uppercase{T}he
  trace-norm distance $D[\rho_1,\rho_2]:=\frac{1}{2}\|\rho_1-\rho_2\|_1$, where
  $\|A\|_1:=\text{Tr}\sqrt{A^{\dagger}A}$, is a standard distance measure
  between quantum states \cite{Nielsen:book}. When $\rho_i$ is a pure state
  $\rho_i=|\psi_i\rangle\langle\psi_i|$ we replace the corresponding argument
  of $D$ by $|\psi_i\rangle$.}\BibitemShut {Stop}%
\bibitem [{\citenamefont {Jansen}\ \emph {et~al.}(2007)\citenamefont {Jansen},
  \citenamefont {Ruskai},\ and\ \citenamefont {Seiler}}]{Jansen:07}%
  \BibitemOpen
  \bibfield  {author} {\bibinfo {author} {\bibfnamefont {S.}~\bibnamefont
  {Jansen}}, \bibinfo {author} {\bibfnamefont {M.-B.}\ \bibnamefont {Ruskai}},
  \ and\ \bibinfo {author} {\bibfnamefont {R.}~\bibnamefont {Seiler}},\ }\href
  {\doibase doi:10.1063/1.2798382} {\bibfield  {journal} {\bibinfo  {journal}
  {J. Math. Phys.}\ }\textbf {\bibinfo {volume} {48}},\ \bibinfo {pages}
  {102111} (\bibinfo {year} {2007})}\BibitemShut {NoStop}%
\bibitem [{\citenamefont {Lidar}\ \emph {et~al.}(2009)\citenamefont {Lidar},
  \citenamefont {Rezakhani},\ and\ \citenamefont {Hamma}}]{lidar:102106}%
  \BibitemOpen
  \bibfield  {author} {\bibinfo {author} {\bibfnamefont {D.~A.}\ \bibnamefont
  {Lidar}}, \bibinfo {author} {\bibfnamefont {A.~T.}\ \bibnamefont
  {Rezakhani}}, \ and\ \bibinfo {author} {\bibfnamefont {A.}~\bibnamefont
  {Hamma}},\ }\href {\doibase 10.1063/1.3236685} {\bibfield  {journal}
  {\bibinfo  {journal} {J. Math. Phys.}\ }\textbf {\bibinfo {volume} {50}},\
  \bibinfo {eid} {102106} (\bibinfo {year} {2009})}\BibitemShut {NoStop}%
\bibitem [{\citenamefont {Wiebe}\ and\ \citenamefont
  {Babcock}(2012)}]{Wiebe:12}%
  \BibitemOpen
  \bibfield  {author} {\bibinfo {author} {\bibfnamefont {N.}~\bibnamefont
  {Wiebe}}\ and\ \bibinfo {author} {\bibfnamefont {N.~S.}\ \bibnamefont
  {Babcock}},\ }\href {\doibase 10.1088/1367-2630/14/1/013024} {\ \textbf
  {\bibinfo {volume} {14}},\ \bibinfo {pages} {013024} (\bibinfo {year}
  {2012})}\BibitemShut {NoStop}%
\bibitem [{\citenamefont {Roland}\ and\ \citenamefont
  {Cerf}(2002)}]{GroverRC:02}%
  \BibitemOpen
  \bibfield  {author} {\bibinfo {author} {\bibfnamefont {J.}~\bibnamefont
  {Roland}}\ and\ \bibinfo {author} {\bibfnamefont {N.}~\bibnamefont {Cerf}},\
  }\href@noop {} {\bibfield  {journal} {\bibinfo  {journal} {Phys. Rev. A}\
  }\textbf {\bibinfo {volume} {65}},\ \bibinfo {pages} {042308} (\bibinfo
  {year} {2002})}\BibitemShut {NoStop}%
\bibitem [{\citenamefont {Rezakhani}\ \emph {et~al.}(2009)\citenamefont
  {Rezakhani}, \citenamefont {Kuo}, \citenamefont {Hamma}, \citenamefont
  {Lidar},\ and\ \citenamefont {Zanardi}}]{RKHLZ:09}%
  \BibitemOpen
  \bibfield  {author} {\bibinfo {author} {\bibfnamefont {A.~T.}\ \bibnamefont
  {Rezakhani}}, \bibinfo {author} {\bibfnamefont {W.-J.}\ \bibnamefont {Kuo}},
  \bibinfo {author} {\bibfnamefont {A.}~\bibnamefont {Hamma}}, \bibinfo
  {author} {\bibfnamefont {D.~A.}\ \bibnamefont {Lidar}}, \ and\ \bibinfo
  {author} {\bibfnamefont {P.}~\bibnamefont {Zanardi}},\ }\href {\doibase
  10.1103/PhysRevLett.103.080502} {\bibfield  {journal} {\bibinfo  {journal}
  {Phys. Rev. Lett.}\ }\textbf {\bibinfo {volume} {103}},\ \bibinfo {pages}
  {080502} (\bibinfo {year} {2009})}\BibitemShut {NoStop}%
\bibitem [{\citenamefont {Brif}\ \emph {et~al.}(2014)\citenamefont {Brif},
  \citenamefont {Grace}, \citenamefont {Sarovar},\ and\ \citenamefont
  {Young}}]{BrifAQC:14}%
  \BibitemOpen
  \bibfield  {author} {\bibinfo {author} {\bibfnamefont {C.}~\bibnamefont
  {Brif}}, \bibinfo {author} {\bibfnamefont {M.~D.}\ \bibnamefont {Grace}},
  \bibinfo {author} {\bibfnamefont {M.}~\bibnamefont {Sarovar}}, \ and\
  \bibinfo {author} {\bibfnamefont {K.~C.}\ \bibnamefont {Young}},\ }\href
  {http://stacks.iop.org/1367-2630/16/i=6/a=065013} {\bibfield  {journal}
  {\bibinfo  {journal} {New Journal of Physics}\ }\textbf {\bibinfo {volume}
  {16}},\ \bibinfo {pages} {065013} (\bibinfo {year} {2014})}\BibitemShut
  {NoStop}%
\bibitem [{\citenamefont {Zeng}\ \emph {et~al.}(2016)\citenamefont {Zeng},
  \citenamefont {Zhang},\ and\ \citenamefont {Sarovar}}]{ZengPathOpt:16}%
  \BibitemOpen
  \bibfield  {author} {\bibinfo {author} {\bibfnamefont {L.}~\bibnamefont
  {Zeng}}, \bibinfo {author} {\bibfnamefont {J.}~\bibnamefont {Zhang}}, \ and\
  \bibinfo {author} {\bibfnamefont {M.}~\bibnamefont {Sarovar}},\ }\href
  {http://stacks.iop.org/1751-8121/49/i=16/a=165305} {\bibfield  {journal}
  {\bibinfo  {journal} {Journal of Physics A: Mathematical and Theoretical}\
  }\textbf {\bibinfo {volume} {49}},\ \bibinfo {pages} {165305} (\bibinfo
  {year} {2016})}\BibitemShut {NoStop}%
\bibitem [{\citenamefont {Johnson}\ \emph {et~al.}(2011)\citenamefont
  {Johnson}, \citenamefont {Amin}, \citenamefont {Gildert}, \citenamefont
  {Lanting}, \citenamefont {Hamze}, \citenamefont {Dickson}, \citenamefont
  {Harris}, \citenamefont {Berkley}, \citenamefont {Johansson}, \citenamefont
  {Bunyk}, \citenamefont {Chapple}, \citenamefont {Enderud}, \citenamefont
  {Hilton}, \citenamefont {Karimi}, \citenamefont {Ladizinsky}, \citenamefont
  {Ladizinsky}, \citenamefont {Oh}, \citenamefont {Perminov}, \citenamefont
  {Rich}, \citenamefont {Thom}, \citenamefont {Tolkacheva}, \citenamefont
  {Truncik}, \citenamefont {Uchaikin}, \citenamefont {Wang}, \citenamefont
  {Wilson},\ and\ \citenamefont {Rose}}]{Dwave}%
  \BibitemOpen
  \bibfield  {author} {\bibinfo {author} {\bibfnamefont {M.~W.}\ \bibnamefont
  {Johnson}}, \bibinfo {author} {\bibfnamefont {M.~H.~S.}\ \bibnamefont
  {Amin}}, \bibinfo {author} {\bibfnamefont {S.}~\bibnamefont {Gildert}},
  \bibinfo {author} {\bibfnamefont {T.}~\bibnamefont {Lanting}}, \bibinfo
  {author} {\bibfnamefont {F.}~\bibnamefont {Hamze}}, \bibinfo {author}
  {\bibfnamefont {N.}~\bibnamefont {Dickson}}, \bibinfo {author} {\bibfnamefont
  {R.}~\bibnamefont {Harris}}, \bibinfo {author} {\bibfnamefont {A.~J.}\
  \bibnamefont {Berkley}}, \bibinfo {author} {\bibfnamefont {J.}~\bibnamefont
  {Johansson}}, \bibinfo {author} {\bibfnamefont {P.}~\bibnamefont {Bunyk}},
  \bibinfo {author} {\bibfnamefont {E.~M.}\ \bibnamefont {Chapple}}, \bibinfo
  {author} {\bibfnamefont {C.}~\bibnamefont {Enderud}}, \bibinfo {author}
  {\bibfnamefont {J.~P.}\ \bibnamefont {Hilton}}, \bibinfo {author}
  {\bibfnamefont {K.}~\bibnamefont {Karimi}}, \bibinfo {author} {\bibfnamefont
  {E.}~\bibnamefont {Ladizinsky}}, \bibinfo {author} {\bibfnamefont
  {N.}~\bibnamefont {Ladizinsky}}, \bibinfo {author} {\bibfnamefont
  {T.}~\bibnamefont {Oh}}, \bibinfo {author} {\bibfnamefont {I.}~\bibnamefont
  {Perminov}}, \bibinfo {author} {\bibfnamefont {C.}~\bibnamefont {Rich}},
  \bibinfo {author} {\bibfnamefont {M.~C.}\ \bibnamefont {Thom}}, \bibinfo
  {author} {\bibfnamefont {E.}~\bibnamefont {Tolkacheva}}, \bibinfo {author}
  {\bibfnamefont {C.~J.~S.}\ \bibnamefont {Truncik}}, \bibinfo {author}
  {\bibfnamefont {S.}~\bibnamefont {Uchaikin}}, \bibinfo {author}
  {\bibfnamefont {J.}~\bibnamefont {Wang}}, \bibinfo {author} {\bibfnamefont
  {B.}~\bibnamefont {Wilson}}, \ and\ \bibinfo {author} {\bibfnamefont
  {G.}~\bibnamefont {Rose}},\ }\href {\doibase 10.1038/nature10012} {\bibfield
  {journal} {\bibinfo  {journal} {Nature}\ }\textbf {\bibinfo {volume} {473}},\
  \bibinfo {pages} {194} (\bibinfo {year} {2011})}\BibitemShut {NoStop}%
\bibitem [{\citenamefont {Lucarelli}(2016)}]{Lucarelli:16}%
  \BibitemOpen
  \bibfield  {author} {\bibinfo {author} {\bibfnamefont {D.}~\bibnamefont
  {Lucarelli}},\ }\href {https://arxiv.org/abs/1611.00188} {\  (\bibinfo {year}
  {2016})},\ \Eprint {http://arxiv.org/abs/ArXiv:1611.00188} {ArXiv:1611.00188}
  \BibitemShut {NoStop}%
\bibitem [{com({\natexlab{b}})}]{comment-energy}%
  \BibitemOpen
  \href@noop {} {}  \bibinfo {note} {\uppercase{A}lternative
  objective functions based on the average energy have been discussed as a
  means for optimizing control schedules for quantum annealing. In
  Ref.~\cite{NehrkornPath:11}, the energy difference at each point in time is
  employed as a figure of merit, while in Ref.~\cite{HerrPath:17}, the average
  energy appears via the fluctuation-dissipation theorem. In both studies, the
  control schedules rely on information about the system at times $t\in[0,T]$.
  CLOAQC relaxes this constraint, requiring only an estimation of the energy of
  the system at $t=T$.}\BibitemShut {Stop}%
\bibitem [{\citenamefont {Spall}(1992)}]{SpallSPSA:92}%
  \BibitemOpen
  \bibfield  {author} {\bibinfo {author} {\bibfnamefont {J.~C.}\ \bibnamefont
  {Spall}},\ }\href {\doibase 10.1109/9.119632} {\bibfield  {journal} {\bibinfo
   {journal} {IEEE Transactions on Automatic Control}\ }\textbf {\bibinfo
  {volume} {37}},\ \bibinfo {pages} {332} (\bibinfo {year} {1992})}\BibitemShut
  {NoStop}%
\bibitem [{\citenamefont {Ferrie}(2014)}]{FerrieSGQT:14}%
  \BibitemOpen
  \bibfield  {author} {\bibinfo {author} {\bibfnamefont {C.}~\bibnamefont
  {Ferrie}},\ }\href {\doibase 10.1103/PhysRevLett.113.190404} {\bibfield
  {journal} {\bibinfo  {journal} {Phys. Rev. Lett.}\ }\textbf {\bibinfo
  {volume} {113}},\ \bibinfo {pages} {190404} (\bibinfo {year}
  {2014})}\BibitemShut {NoStop}%
\bibitem [{\citenamefont {Ferrie}\ and\ \citenamefont
  {Moussa}(2015)}]{FerrieQC:15}%
  \BibitemOpen
  \bibfield  {author} {\bibinfo {author} {\bibfnamefont {C.}~\bibnamefont
  {Ferrie}}\ and\ \bibinfo {author} {\bibfnamefont {O.}~\bibnamefont
  {Moussa}},\ }\href {\doibase 10.1103/PhysRevA.91.052306} {\bibfield
  {journal} {\bibinfo  {journal} {Phys. Rev. A}\ }\textbf {\bibinfo {volume}
  {91}},\ \bibinfo {pages} {052306} (\bibinfo {year} {2015})}\BibitemShut
  {NoStop}%
\bibitem [{\citenamefont {Granade}(2015)}]{Granade:15}%
  \BibitemOpen
  \bibfield  {author} {\bibinfo {author} {\bibfnamefont {C.~E.}\ \bibnamefont
  {Granade}},\ }\emph {\bibinfo {title} {Characterization, Verification, and
  Control for Large Quantum Systems}},\ \href@noop {} {Ph.D. thesis} (\bibinfo
  {year} {2015})\BibitemShut {NoStop}%
\bibitem [{\citenamefont {Spall}(1998)}]{SpallSPSAImp:98}%
  \BibitemOpen
  \bibfield  {author} {\bibinfo {author} {\bibfnamefont {J.~C.}\ \bibnamefont
  {Spall}},\ }\href {\doibase 10.1109/7.705889} {\bibfield  {journal} {\bibinfo
   {journal} {IEEE Transactions on Aerospace and Electronic Systems}\ }\textbf
  {\bibinfo {volume} {34}},\ \bibinfo {pages} {817} (\bibinfo {year}
  {1998})}\BibitemShut {NoStop}%
\bibitem [{\citenamefont {Sadegh}\ and\ \citenamefont
  {Spall}(1998)}]{SadeghSPSA:98}%
  \BibitemOpen
  \bibfield  {author} {\bibinfo {author} {\bibfnamefont {P.}~\bibnamefont
  {Sadegh}}\ and\ \bibinfo {author} {\bibfnamefont {J.~C.}\ \bibnamefont
  {Spall}},\ }\href {\doibase 10.1109/9.720513} {\bibfield  {journal} {\bibinfo
   {journal} {IEEE Transactions on Automatic Control}\ }\textbf {\bibinfo
  {volume} {43}},\ \bibinfo {pages} {1480} (\bibinfo {year}
  {1998})}\BibitemShut {NoStop}%
\bibitem [{\citenamefont {Grover}(1997)}]{Grover:97}%
  \BibitemOpen
  \bibfield  {author} {\bibinfo {author} {\bibfnamefont {L.~K.}\ \bibnamefont
  {Grover}},\ }\href {\doibase 10.1103/PhysRevLett.79.325} {\bibfield
  {journal} {\bibinfo  {journal} {Phys. Rev. Lett.}\ }\textbf {\bibinfo
  {volume} {79}},\ \bibinfo {pages} {325} (\bibinfo {year} {1997})}\BibitemShut
  {NoStop}%
\bibitem [{\citenamefont {Bennett}\ \emph {et~al.}(1997)\citenamefont
  {Bennett}, \citenamefont {Bernstein}, \citenamefont {Brassard},\ and\
  \citenamefont {Vazirani}}]{Bennett:97}%
  \BibitemOpen
  \bibfield  {author} {\bibinfo {author} {\bibfnamefont {C.~H.}\ \bibnamefont
  {Bennett}}, \bibinfo {author} {\bibfnamefont {E.}~\bibnamefont {Bernstein}},
  \bibinfo {author} {\bibfnamefont {G.}~\bibnamefont {Brassard}}, \ and\
  \bibinfo {author} {\bibfnamefont {U.}~\bibnamefont {Vazirani}},\ }\href
  {\doibase 10.1137/S0097539796300933} {\bibfield  {journal} {\bibinfo
  {journal} {SIAM Journal on Computing}\ }\textbf {\bibinfo {volume} {26}},\
  \bibinfo {pages} {1510} (\bibinfo {year} {1997})},\ \Eprint
  {http://arxiv.org/abs/https://doi.org/10.1137/S0097539796300933}
  {https://doi.org/10.1137/S0097539796300933} \BibitemShut {NoStop}%
\bibitem [{\citenamefont {Santra}\ \emph {et~al.}(2014)\citenamefont {Santra},
  \citenamefont {Quiroz}, \citenamefont {Steeg},\ and\ \citenamefont
  {Lidar}}]{Santra2SAT:14}%
  \BibitemOpen
  \bibfield  {author} {\bibinfo {author} {\bibfnamefont {S.}~\bibnamefont
  {Santra}}, \bibinfo {author} {\bibfnamefont {G.}~\bibnamefont {Quiroz}},
  \bibinfo {author} {\bibfnamefont {G.~V.}\ \bibnamefont {Steeg}}, \ and\
  \bibinfo {author} {\bibfnamefont {D.~A.}\ \bibnamefont {Lidar}},\ }\href
  {http://stacks.iop.org/1367-2630/16/i=4/a=045006} {\bibfield  {journal}
  {\bibinfo  {journal} {New Journal of Physics}\ }\textbf {\bibinfo {volume}
  {16}},\ \bibinfo {pages} {045006} (\bibinfo {year} {2014})}\BibitemShut
  {NoStop}%
\bibitem [{\citenamefont {Coppersmith}\ \emph {et~al.}(2004)\citenamefont
  {Coppersmith}, \citenamefont {Gamarnik}, \citenamefont {Hajiaghayi},\ and\
  \citenamefont {Sorkin}}]{Coppersmith2SAT:04}%
  \BibitemOpen
  \bibfield  {author} {\bibinfo {author} {\bibfnamefont {D.}~\bibnamefont
  {Coppersmith}}, \bibinfo {author} {\bibfnamefont {D.}~\bibnamefont
  {Gamarnik}}, \bibinfo {author} {\bibfnamefont {M.}~\bibnamefont
  {Hajiaghayi}}, \ and\ \bibinfo {author} {\bibfnamefont {G.~B.}\ \bibnamefont
  {Sorkin}},\ }\href {\doibase 10.1002/rsa.20015} {\bibfield  {journal}
  {\bibinfo  {journal} {Random Structures and Algorithms}\ }\textbf {\bibinfo
  {volume} {24}},\ \bibinfo {pages} {502} (\bibinfo {year} {2004})}\BibitemShut
  {NoStop}%
\bibitem [{\citenamefont {Crosson}\ \emph {et~al.}(2014)\citenamefont
  {Crosson}, \citenamefont {Farhi}, \citenamefont {Yen-Yu~Lin}, \citenamefont
  {Lin},\ and\ \citenamefont {Shor}}]{CrossonNonSto:14}%
  \BibitemOpen
  \bibfield  {author} {\bibinfo {author} {\bibfnamefont {E.}~\bibnamefont
  {Crosson}}, \bibinfo {author} {\bibfnamefont {E.}~\bibnamefont {Farhi}},
  \bibinfo {author} {\bibfnamefont {C.}~\bibnamefont {Yen-Yu~Lin}}, \bibinfo
  {author} {\bibfnamefont {H.-H.}\ \bibnamefont {Lin}}, \ and\ \bibinfo
  {author} {\bibfnamefont {P.}~\bibnamefont {Shor}},\ }\href
  {https://arxiv.org/abs/1401.7320} {\  (\bibinfo {year} {2014})},\ \Eprint
  {http://arxiv.org/abs/ArXiv: 1401.7320} {ArXiv: 1401.7320} \BibitemShut
  {NoStop}%
\bibitem [{\citenamefont {Farhi}\ \emph {et~al.}(2011)\citenamefont {Farhi},
  \citenamefont {Goldstone},\ and\ \citenamefont {Gutmann}}]{FarhiNonSto:11}%
  \BibitemOpen
  \bibfield  {author} {\bibinfo {author} {\bibfnamefont {E.}~\bibnamefont
  {Farhi}}, \bibinfo {author} {\bibfnamefont {J.}~\bibnamefont {Goldstone}}, \
  and\ \bibinfo {author} {\bibfnamefont {S.}~\bibnamefont {Gutmann}},\ }\href
  {https://arxiv.org/abs/quant-ph/0208135} {\  (\bibinfo {year} {2011})},\
  \Eprint {http://arxiv.org/abs/quant-ph/0208135} {quant-ph/0208135}
  \BibitemShut {NoStop}%
\bibitem [{\citenamefont {Hormozi}\ \emph {et~al.}(2017)\citenamefont
  {Hormozi}, \citenamefont {Brown}, \citenamefont {Carleo},\ and\ \citenamefont
  {Troyer}}]{HormoziNonSto:17}%
  \BibitemOpen
  \bibfield  {author} {\bibinfo {author} {\bibfnamefont {L.}~\bibnamefont
  {Hormozi}}, \bibinfo {author} {\bibfnamefont {E.~W.}\ \bibnamefont {Brown}},
  \bibinfo {author} {\bibfnamefont {G.}~\bibnamefont {Carleo}}, \ and\ \bibinfo
  {author} {\bibfnamefont {M.}~\bibnamefont {Troyer}},\ }\href {\doibase
  10.1103/PhysRevB.95.184416} {\bibfield  {journal} {\bibinfo  {journal} {Phys.
  Rev. B}\ }\textbf {\bibinfo {volume} {95}},\ \bibinfo {pages} {184416}
  (\bibinfo {year} {2017})}\BibitemShut {NoStop}%
\bibitem [{\citenamefont {Nishimori}\ and\ \citenamefont
  {Takada}(2017)}]{NishimoriNonSto:17}%
  \BibitemOpen
  \bibfield  {author} {\bibinfo {author} {\bibfnamefont {H.}~\bibnamefont
  {Nishimori}}\ and\ \bibinfo {author} {\bibfnamefont {K.}~\bibnamefont
  {Takada}},\ }\href {\doibase 10.3389/fict.2017.00002} {\bibfield  {journal}
  {\bibinfo  {journal} {Frontiers in ICT}\ }\textbf {\bibinfo {volume} {4}},\
  \bibinfo {pages} {2} (\bibinfo {year} {2017})}\BibitemShut {NoStop}%
\bibitem [{\citenamefont {Childs}\ \emph {et~al.}(2001)\citenamefont {Childs},
  \citenamefont {Farhi},\ and\ \citenamefont {Preskill}}]{ChildsAQC:01}%
  \BibitemOpen
  \bibfield  {author} {\bibinfo {author} {\bibfnamefont {A.~M.}\ \bibnamefont
  {Childs}}, \bibinfo {author} {\bibfnamefont {E.}~\bibnamefont {Farhi}}, \
  and\ \bibinfo {author} {\bibfnamefont {J.}~\bibnamefont {Preskill}},\ }\href
  {\doibase 10.1103/PhysRevA.65.012322} {\bibfield  {journal} {\bibinfo
  {journal} {Phys. Rev. A}\ }\textbf {\bibinfo {volume} {65}},\ \bibinfo
  {pages} {012322} (\bibinfo {year} {2001})}\BibitemShut {NoStop}%
\bibitem [{\citenamefont {Nielsen}\ and\ \citenamefont
  {Chuang}(2000)}]{Nielsen:book}%
  \BibitemOpen
  \bibfield  {author} {\bibinfo {author} {\bibfnamefont {M.~A.}\ \bibnamefont
  {Nielsen}}\ and\ \bibinfo {author} {\bibfnamefont {I.~L.}\ \bibnamefont
  {Chuang}},\ }\href@noop {} {\emph {\bibinfo {title} {Quantum Computation and
  Quantum Information}}}\ (\bibinfo  {publisher} {Cambridge University Press},\
  \bibinfo {address} {New York},\ \bibinfo {year} {2000})\BibitemShut {NoStop}%
\bibitem [{\citenamefont {Nehrkorn}\ \emph {et~al.}(2011)\citenamefont
  {Nehrkorn}, \citenamefont {Montangero}, \citenamefont {Ekert}, \citenamefont
  {Smerzi}, \citenamefont {Fazio},\ and\ \citenamefont
  {Calarco}}]{NehrkornPath:11}%
  \BibitemOpen
  \bibfield  {author} {\bibinfo {author} {\bibfnamefont {J.}~\bibnamefont
  {Nehrkorn}}, \bibinfo {author} {\bibfnamefont {S.}~\bibnamefont
  {Montangero}}, \bibinfo {author} {\bibfnamefont {A.}~\bibnamefont {Ekert}},
  \bibinfo {author} {\bibfnamefont {A.}~\bibnamefont {Smerzi}}, \bibinfo
  {author} {\bibfnamefont {R.}~\bibnamefont {Fazio}}, \ and\ \bibinfo {author}
  {\bibfnamefont {T.}~\bibnamefont {Calarco}},\ }\href
  {https://arxiv.org/abs/1105.1707} {\  (\bibinfo {year} {2011})},\ \Eprint
  {http://arxiv.org/abs/ArXiv:1105.1707} {ArXiv:1105.1707} \BibitemShut
  {NoStop}%
\bibitem [{\citenamefont {Herr}\ \emph {et~al.}(2017)\citenamefont {Herr},
  \citenamefont {Brown}, \citenamefont {Heim}, \citenamefont {Konz},
  \citenamefont {Mazzola},\ and\ \citenamefont {Troyer}}]{HerrPath:17}%
  \BibitemOpen
  \bibfield  {author} {\bibinfo {author} {\bibfnamefont {D.}~\bibnamefont
  {Herr}}, \bibinfo {author} {\bibfnamefont {E.}~\bibnamefont {Brown}},
  \bibinfo {author} {\bibfnamefont {B.}~\bibnamefont {Heim}}, \bibinfo {author}
  {\bibfnamefont {M.}~\bibnamefont {Konz}}, \bibinfo {author} {\bibfnamefont
  {G.}~\bibnamefont {Mazzola}}, \ and\ \bibinfo {author} {\bibfnamefont
  {M.}~\bibnamefont {Troyer}},\ }\href {https://arxiv.org/abs/1705.00420} {\
  (\bibinfo {year} {2017})},\ \Eprint {http://arxiv.org/abs/ArXiv:1705.00420}
  {ArXiv:1705.00420} \BibitemShut {NoStop}%
\bibitem [{\citenamefont {Takahashi}(2017)}]{TakahashiAQC:17}%
  \BibitemOpen
  \bibfield  {author} {\bibinfo {author} {\bibfnamefont {K.}~\bibnamefont
  {Takahashi}},\ }\href {\doibase 10.1103/PhysRevA.95.012309} {\bibfield
  {journal} {\bibinfo  {journal} {Phys. Rev. A}\ }\textbf {\bibinfo {volume}
  {95}},\ \bibinfo {pages} {012309} (\bibinfo {year} {2017})}\BibitemShut
  {NoStop}%
\end{thebibliography}
%

%
%

\newpage
\appendix

\section{Runtime discussion}
\label{sec:app-runtime}
The runtime $T$ is a free parameter that must be selected at the start of a CLOAQC optimization. While the adiabatic theorem provides some guidance for selecting an appropriate $T$, potential uncertainties in $H$ and computational demands required to numerically approximate the minimum spectral gap $\Delta$ typically hinder one's ability to accurately estimate the lower bound on $T$. Therefore, selecting a $T$ which satisfies the adiabatic theorem and maintains high ground state probability is a challenging task in both the closed and open quantum system setting. Since CLOAQC relies on the sampling of the ground state of $H_P$ to successfully optimize the control schedules, one would expect the protocol's performance to highly dependent upon the choice of $T$. Below, the dependence of CLOAQC performance on $T$ is investigated. CLOAQC is shown to be substantially improve the probability of sampling the ground state of $H_P$ even in cases where the ground state probability is low at the initialization of the protocol.

\begin{figure}[t]
\centering
\includegraphics[width=0.8\columnwidth]{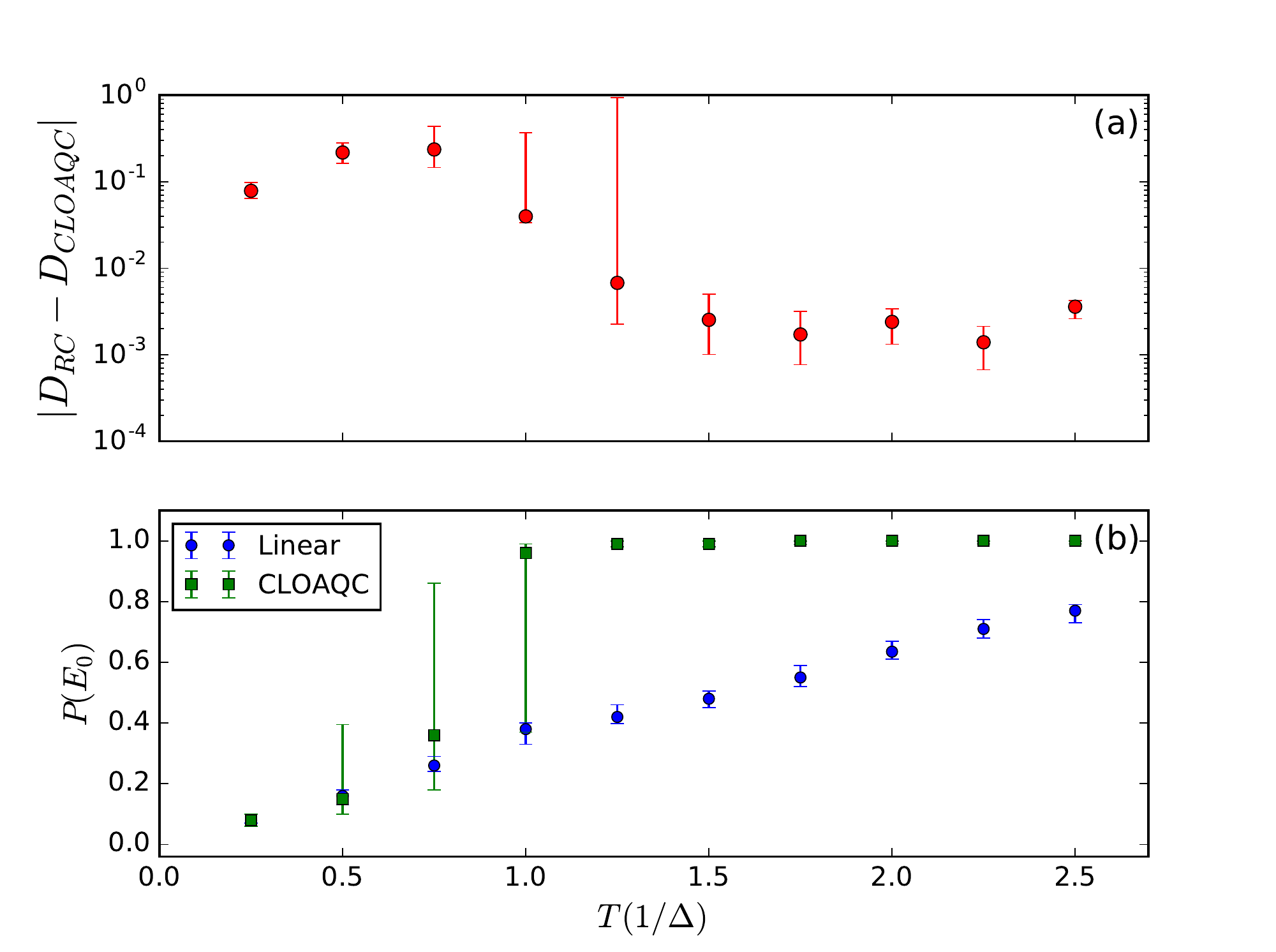}
\caption{CLOAQC performance as a function of the adiabatic runtime $T$ for GSA using one independent control. Panel (a): Relative difference in adiabatic error between CLOAQC optimized path and RC solution. Panel (b): Probability of sampling the ground state for the initial (linear) control schedule and CLOAQC. Distributions include 100 realizations of CLOAQC with markers and whiskers denoted medians and interquartile ranges, respectively. CLOAQC is implemented using $K=5000$ and $M=100$.}
\label{fig:gsa-ta}
\end{figure}

\begin{figure}[t]
\centering
\includegraphics[width=0.8\columnwidth]{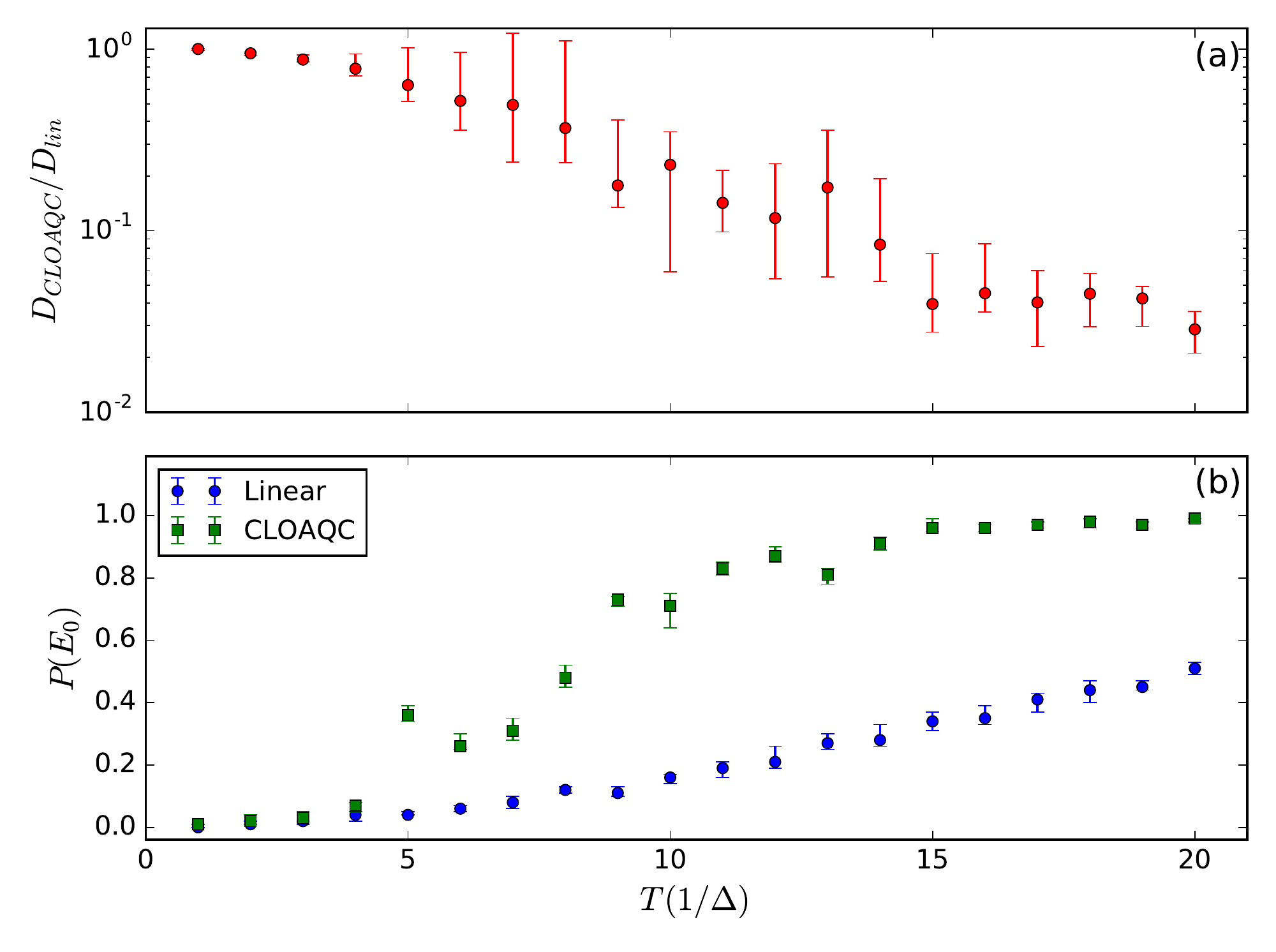}
\caption{CLOAQC performance as a function of the adiabatic runtime $T$ for one representative MAX 2-SAT USA instance using one independent control. Panel (a): Ratio of adiabatic errors for the CLOAQC optimized path and the linear control schedule. Panel (b): Probability of sampling the ground state for the initial (linear) control schedule and CLOAQC. Distributions include 25 realizations of CLOAQC with markers and whiskers denoted medians and interquartile ranges, respectively. CLOAQC is implemented using $K=1000$ and $M=100$.}
\label{fig:max2sat-ta}
\end{figure}

Performance comparisons as a function of $T$ are presented for both GSA and MAX 2-SAT. In Figure~\ref{fig:gsa-ta}, CLOAQC performance is assessed as a function of $T$ for GSA for one independent control. CLOAQC optimizations are performed with $K=5000$ and $M=100$ for 100 realizations of the protocol. The relative difference between CLOAQC adiabatic error and the RC solution adiabatic error is displayed in panel (a) along with the corresponding probability of obtaining the ground state energy $E_0$ for both the initial (linear) schedule and the CLOAQC optimized control schedule in panel (b). In Figure~\ref{fig:max2sat-ta}, the ratio of adiabatic errors and frequency of $E_0$ for the linear and CLOAQC optimized controls are compared in panels (a) and (b), respectively, for the MAX 2-SAT problem with one independent control. CLOAQC is implemented with $K=1000$ and $M=100$ for 25 realizations. Figure~\ref{fig:max2sat-ta} focuses on one USA instance explored in the main text, however, the results shown here capture the typical behavior for the remaining 99 instances considered. Note that for both the GSA and MAX 2-SAT results, markers denote medians, while wiskers denote the interquartile range.

CLOAQC successfully optimize control schedules when initialized with controls paths that yield relatively low probabilities of sampling the ground state of $H_P$. As $T$ increases, CLOAQC is able to more accurately reproduce the optimal RC path for GSA and offer improvements in adiabatic error over the linear schedule for MAX 2-SAT. This observation is evident from both the trace norm distance and the probability of sampling $E_0$, where $E_0$ is sampled with unit median probability for the CLOAQC optimized path at $T\approx 1.75/\Delta$ and $T\approx 20/\Delta$ for GSA and MAX 2-SAT, respectively. Note that in each case, the initial, linear control schedule only samples $E_0$ with probability $P(E_0)\approx 50\%$. Thus, CLOAQC doubles the probability of sampling the ground state. Substantial improvements in $P(E_0)$ are also observed for shorter $T$ values. For example, for GSA, CLOAQC can achieve a median $P(E_0)\approx 96\%$ at $T\approx 1/\Delta$, where the linear control schedule yields $P(E_0)\approx 38\%$. Similarly, for MAX-2SAT, at $T\approx 15/\Delta$, CLOAQC achieves $P(E_0)\approx 96\%$ using a linear schedule with $P(E_0)\approx 35\%-40\%$ for all problem instances considered. At $T\approx 5/\Delta$, a linear schedule with $P(E_0)\approx 2\%-7\%$ ultimately yields an optimized $P(E_0)\approx 30\%-39\%$. Lastly, it is important to note that CLOAQC is also capable of attaining non-zero $P(E_0)$ when initialized at $T$ values where $P(E_0)=0$. Such is the case at $T\approx 1/\Delta$ for a majority of the MAX 2-SAT USA instances discussed here.

\section{Alternative Intermediate Hamiltonians}
\label{sec:app-hi}
\begin{figure}[t]
\includegraphics[width=0.9\columnwidth]{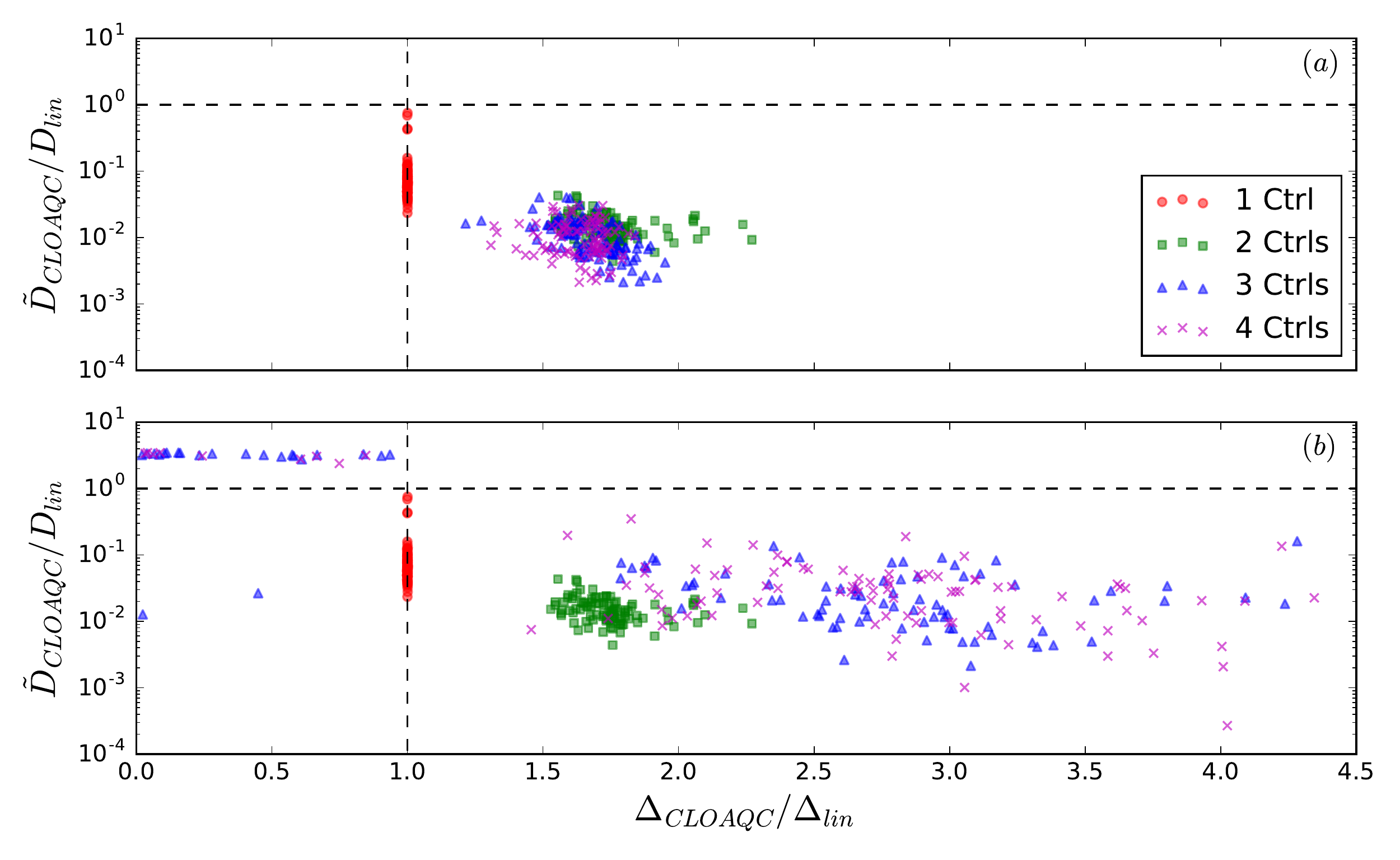}
\caption{Comparison of CLOAQC vs. a linear schedule for 100 instances of the MAX 2-SAT problem for four different control scenarios. Each panel directly corresponds to the two $H_I$ scenarios considered.}
\label{fig:max2sat-alt-hi}
\end{figure}

The benefits of non-stoquastic Hamiltonians combined with optimized controls provided by CLOAQC are further explored here for the MAX 2-SAT problem using alternative Hamiltonians to the two-local XX interaction Hamiltonian presented in the main text. In Figure~\ref{fig:max2sat-alt-hi}(a) and (b), the intermediate Hamiltonian is given by $H^{(y)}_I=\sum_j \sy_j$ and $H^{(xz)}_I=\sum_{ij}\sx_i\sz_j+\sz_i\sx_j$, respectively. The former has been previously studied as an additional driving term for quantum annealing, where techniques from ``shortcuts to adiabaticity" where employed to produce optimized control for particular Ising-type problem Hamiltonians~\cite{TakahashiAQC:17}. A Hamiltonian similar to $H^{(xz)}_I$ has been previously employed as a calatyst Hamiltonian that improves the success probability and runtime scaling for a specific choice of $H_P$~\cite{FarhiNonSto:11}. This study focuses on potential improvements provided by each $H_I$ in conjunction with CLOAQC optimized control for the 100 USA problem instances discussed in the main text.

The addition of $H^{(y)}_I$ leads to improvements in adiabatic error and no substantial improvements in the minimum spectral gap. In Figure~\ref{fig:max2sat-alt-hi}(a), the median performance of CLOAQC is compared to the one control linear ramping schedule. The one and two control schedules are included for reference, with the results for three and four controls constituting the $H_I$-dependent distributions. Significantly more localized than the XX interaction Hamiltonian, $H^{(y)}_I$ with optimized control generally offers improvements in adiabatic error without the need for gap amplification for the 100 USA problems considered here. Degradations in adiabatic error for the four control case are still observed and again attributed to fixing the number of iterations and energy samples used for the CLOAQC optimization experiments. 

In contrast $H^{(y)}$, improvements in adiabatic error and minimum gap size are strongly instance dependent when utilizing $H^{(xz)}_I$. Improvements in adiabatic error up to a factor of approximately $10^{-3}$ and $10^{-4}$ are observed for three and four independent controls, respectively, with gap amplifications up to approximately $4.5\times \Delta_{lin}$. While substantial improvements are observed for a subset of instances, approximately 1/3 of the instances do not benefit from the addition of $H^{(xz)}_I$ with optimized control. Tracking the adiabatic error as a function of iteration, the addition of $H^{(xz)}_I$ appears to induce stability issues in CLOAQC that lead to limited success in optimization over the 25 realizations. By altering the convergence parameters, it is possible to improve median performance and overcome such issues at the cost of an increase in the number of iterations. 

Median improvements in adiabatic errors and gap amplification taken over the distribution of problem instances are summarized in Table~\ref{tbl:max2sat-sum} for all choices of $H_I$.

\begin{table}
\centering
\begin{tabular}{ccccc}
\hline\hline
\multirow{2}{*}{$H_I$} & \multicolumn{4}{c}{\# of Ctrls}\\ \cline{2-5}
& 1 & 2& 3&4\\ \hline
$H^{(xx)}_I$ & (0.065, 1) & (0.014, 1.735) & (0.012, 2.973) & (0.062, 2.538)\\
$H^{(y)}_I$ & SAA & SAA & (0.009, 1.673) & (0.011, 1.658)\\
$H^{(xz)}_I$ & SAA & SAA & (0.023, 2.604) & (0.031, 2.722)\\
\hline\hline
\end{tabular}
\caption{Summary of median improvements in adiabatic error and gap amplifications for the distribution of 100 MAX 2-SAT USA problem instances for each non-stoquastic Hamiltonian considered. Data is formatted as $(\alpha_D, \alpha_\Delta)$, where $\alpha_D=\tilde{D}_{CLOAQC}/D_{lin}$ and the ratio of minimum gaps is given by $\alpha_\Delta=\Delta_{CLOAQC}/\Delta_{lin})$. SAA denotes ``same as above" for equivalent results among different intermediate Hamiltonians.}
\label{tbl:max2sat-sum}
\end{table}


\section{Control Schedules}
\label{sec:app-ctrls}
\subsection{Grover's Search Algorithm}
\begin{figure}[t]
\centering
\includegraphics[width=0.8\columnwidth]{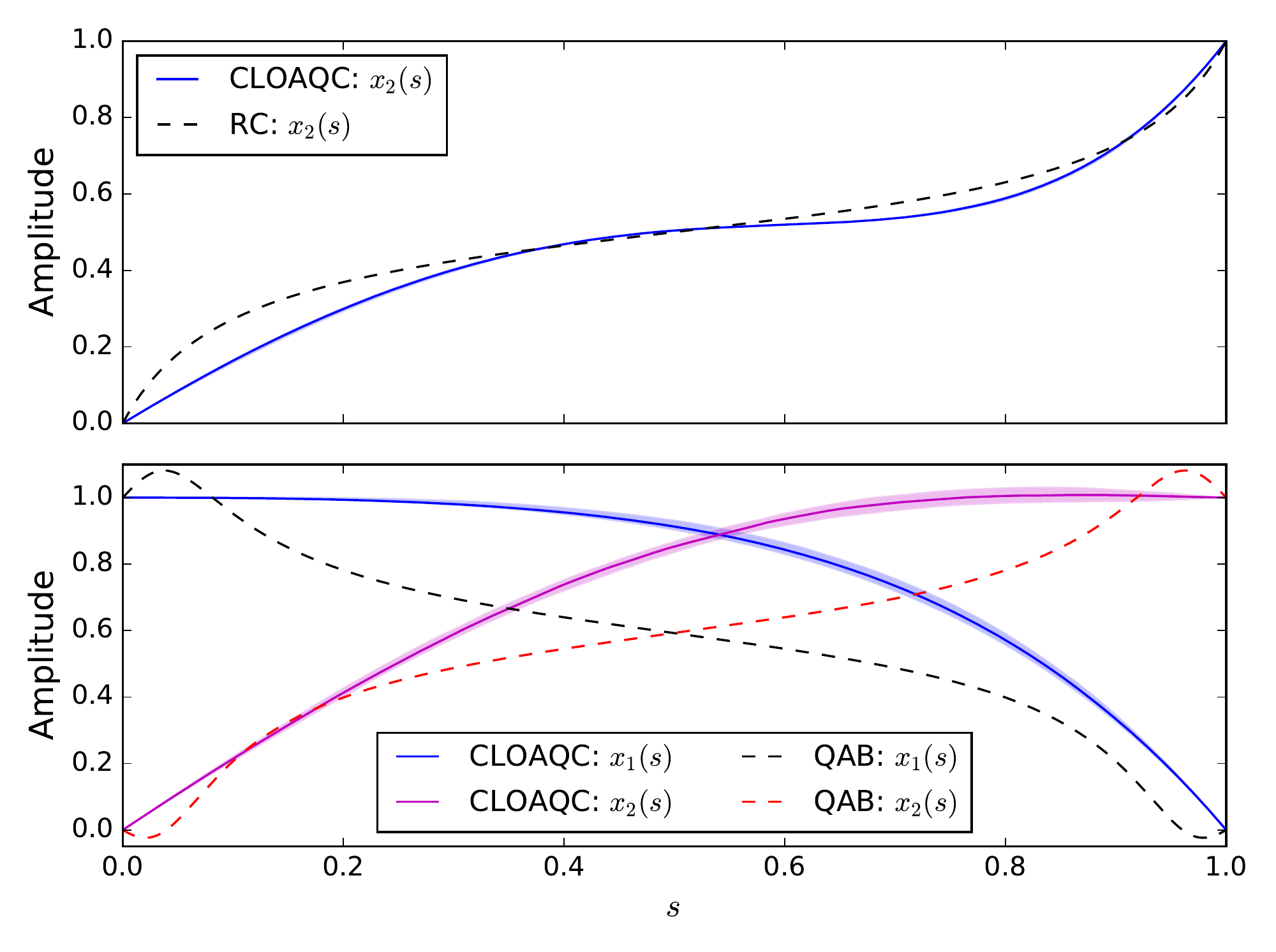}
\caption{Comparison between CLOAQC optimized control paths and RC (top) and QAB (bottom) control paths for $n=4$. CLOAQC is implemented with $K=10000$, and $M=100$. The median paths are shown by dark lines while shaded regions correspond to interquartile range for 100 realizations of CLOAQC. CLOAQC control paths correspond to the results presented in main text (Figure~\ref{fig:grover}). }
\label{fig:gsa-paths}
\end{figure}

A comparison between CLOAQC and the time-optimal GSA solutions corresponding to the main text results (Figure~\ref{fig:grover}) for $n=4$ and $M=100$ is shown in Figure~\ref{fig:gsa-paths}. The one control case yields optimized controls that closely resemble the RC control path, as seen from the top panel of Figure~\ref{fig:gsa-paths}. Deviations from the RC solution are attributed to the truncated polynomial expansion used to define the CLOAQC control field. In contrast, the complexity of the QAB profile is far more difficult for CLOAQC to reproduce under the fifth order polynomial expansion; hence, a greater distinction between the CLOAQC optimized path and the QAB profile is observed; see Figure~\ref{fig:gsa-paths}, bottom panel. Interestingly, even with such a distinction, the relative difference in performance between CLOAQC and the QAB path is roughly only $10^{-3}$ for $n=4$ and less than that for $n=6$ and $n=8$.

\subsection{MAX 2-SAT}
\begin{figure}[t]
\centering
\includegraphics[width=0.8\columnwidth]{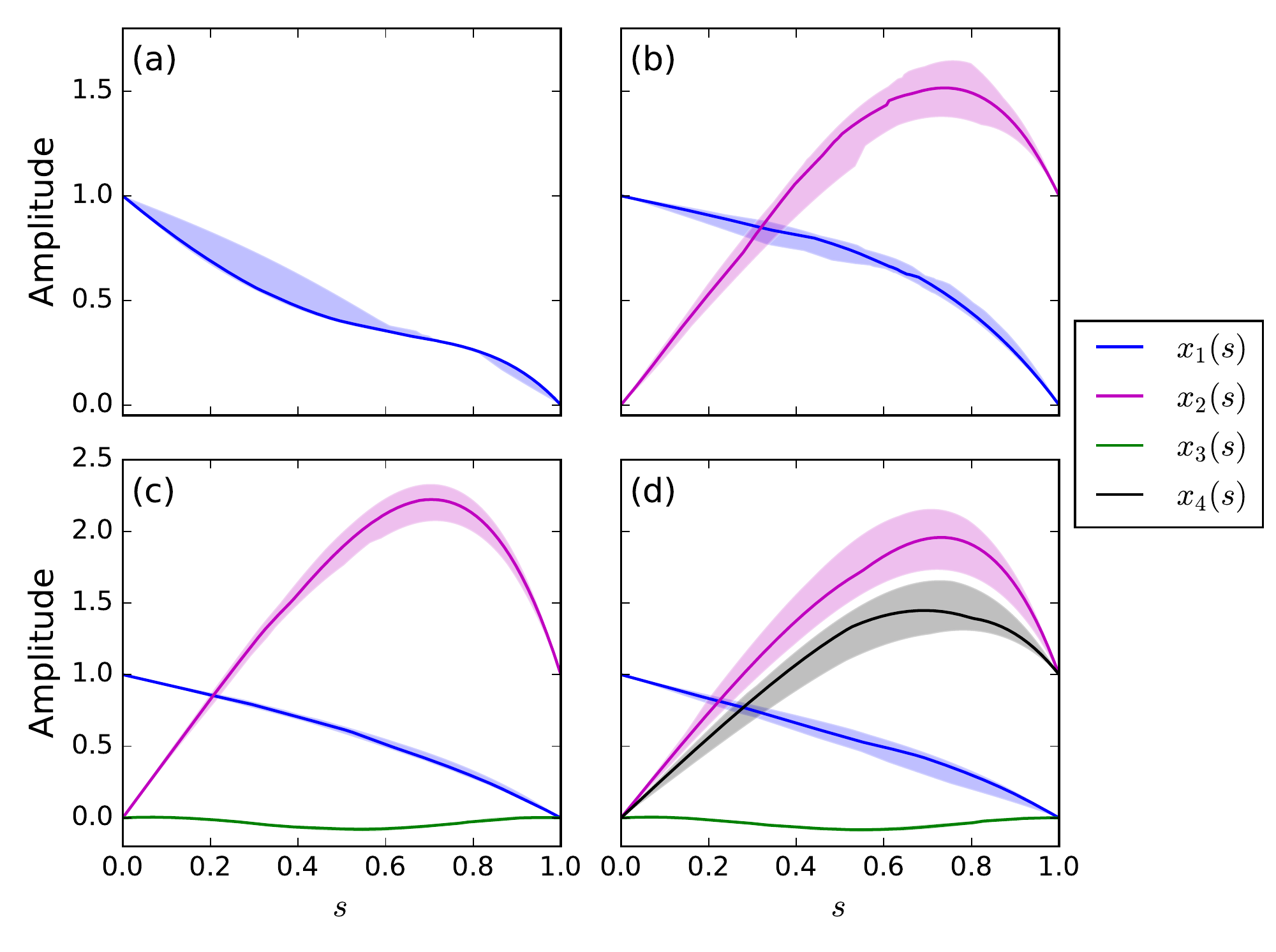}
\caption{CLOAQC control paths resulting from 25 realization of CLOAQC implemented with $K=1000$ and $M=100$. Results for 1-4 independent controls are shown in panels (a)-(d), respectively. The intermediate Hamiltonian is the two-local XX interaction discussed in the main text.}
\label{fig:max2sat-ctrls-xx}
\end{figure}

\begin{figure}[t]
\centering
\includegraphics[width=\columnwidth]{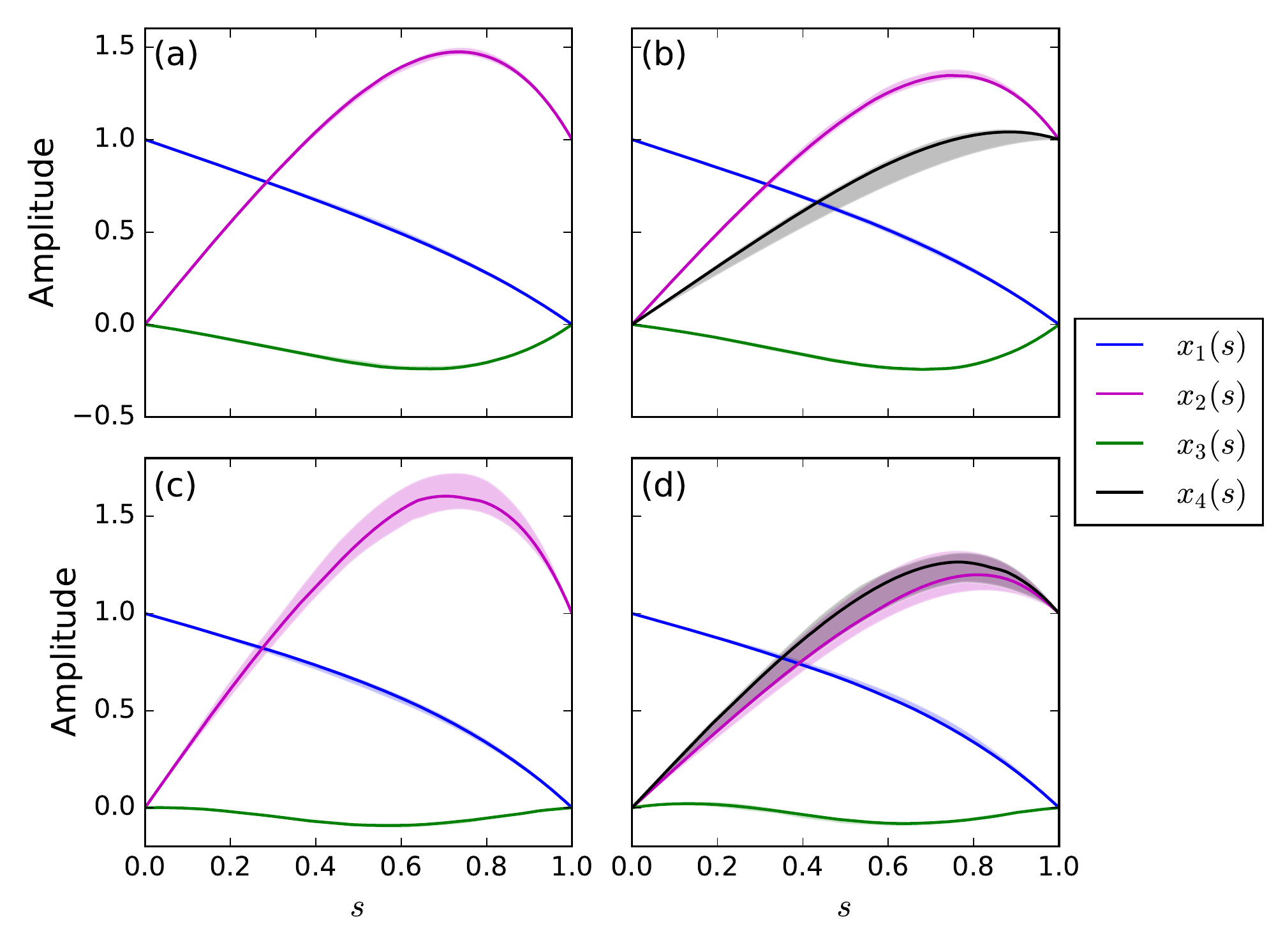}
\caption{CLOAQC control paths resulting from 25 realizations of CLOAQC implemented with $K=1000$ and $M=100$. Panels (a) and (b) correspond to the $H^{(y)}_I$ intermediate Hamiltonian for three and four independent controls, respectively. Panels (c) and (d) correspond to the $H^{(xz)}_I$ intermediate Hamiltonian for three and four independent controls, respectively.}
\label{fig:max2sat-ctrls-yxz}
\end{figure}

The CLOAQC optimized control paths for one MAX 2-SAT instance are shown in Figures~\ref{fig:max2sat-ctrls-xx} and \ref{fig:max2sat-ctrls-yxz}. The main text results for $H^{(xx)}_I$ are shown in Figure~\ref{fig:max2sat-ctrls-xx} for 1-4 independent controls. Lines correspond to the median path, while shaded regions denote the interquartile range for 25 CLOAQC realizations. Figure~\ref{fig:max2sat-ctrls-yxz} displays similar results for $H^{(y)}_I$ and $H^{(xz)}_I$ for three and four controls only. The minimum spectral gap for this particular problem instance lies between $s=0.6$ and $s=0.8$ for a majority of the control schedules. Note that the most significant ramping of the control schedules typically occurs within this range for $x_j(s)$, $j=2,3,4$.

\end{document}